\documentclass[10pt]{article}

\usepackage{cite} 
\usepackage{hyperref}
\usepackage{amsmath,amssymb}
\usepackage{authblk}
\usepackage{graphicx}
\usepackage[shortlabels]{enumitem}
\usepackage{xcolor}

\def\be{\begin{equation}} 
\def\ee{\end{equation}}   
\def \eea{\end{eqnarray}}
\def \bea{\begin{eqnarray}}

\newcommand{\dx}{ \dot{x} }

\newcommand{\Wr}{ \Omega_r }
\newcommand{\Wm}{ \Omega_m }
\newcommand{\Wl}{ \Omega_\Lambda }

\newcommand{\km}{ \text{ km}}
\newcommand{\s}{ \text{ s}}
\newcommand{\Mpc}{ \text{ Mpc}}
\newcommand{\yr}{ \text{ yr}}
\newcommand{\Gyr}{ \text{ Gyr}}

\newcommand{\LCDM}{ \Lambda\text{CDM}}

\newcommand{\uantof}{Departamento de Física, Universidad de Antofagasta, Aptdo. 02800, Chile}

\newcommand{\puc}{Instituto de F{\'i}sica, Pontificia Cat{\'o}lica Universidad de Chile, Av. Vicu{\~n}a Mackenna 4860, Santiago, Chile}
\newcommand{\viena}{Institut f\"ur Theoretische Physik,
 Technische Universit\"at Wien,
 Wiedner Hauptstrasse 8-10,
 A-1040 Vienna, Austria}

\newcommand{\pucv}{Instituto de Física, Pontificia Universidad Católica de Valparaíso,
Avenida Brasil 2950, Casilla 4059, Valparaíso, Chile.}

\begin{document}
\title{Can scale--dependent cosmology alleviate\\ the $H_0$ tension?}
\author[1]{Pedro D. Alvarez \thanks{E-mail: \href{mailto:pedro.alvarez@uantof.cl}{\nolinkurl{pedro.alvarez@uantof.cl}}}}
\affil[1]{\uantof}

\author[2,3]{Benjamin Koch \thanks{E-mail: \href{mailto:benjamin.koch@tuwien.ac.at}{\nolinkurl{benjamin.koch@tuwien.ac.at}}}}
\affil[2]{\viena}
\affil[3]{\puc}

\author[3]{Cristobal Laporte \thanks{E-mail: \href{mailto:calaporte@uc.cl}{\nolinkurl{calaporte@uc.cl}}}}

\author[4]{\'Angel Rinc\'on \thanks{E-mail: \href{mailto:angel.rincon@pucv.cl}{\nolinkurl{angel.rincon@pucv.cl}}}}
\affil[4]{\pucv}

\maketitle

\begin{abstract}
    Scale--dependence is a common feature to all effective models of quantum gravity.
    In this paper, a cosmological model based on the scale--dependent scenario of gravity is presented.
    It is argued that such models, where the scale--dependence appears as a correction to the classical $\Lambda$CDM evolution,
    have the potential of addressing the tensions between early and late time measurements of $H_0$.
    After defining criteria to parametrize this tension, we perform
    a numerical scan over the parameter space of the scale--dependent model, subject to these criteria.
    In this analysis, it is found that, indeed, 
    the tension can be released.
\end{abstract}

\tableofcontents

\section{Introduction}\label{Intro}

We live in an exciting era of cosmology where five out of six cosmological parameters can be measured at a sub percent precision level from properties of the CMB \cite{Aghanim:2018eyx}. This allowed us to set constraints to models beyond $\LCDM$ and models of inflation \cite{Ade:2015rim,DiValentino:2019dzu,Akrami:2018odb}. 

In spite of this tremendous success, it appears to be a tension between the CMB-inferred value and late cosmological time measurements of the Hubble constant at the level of four sigma~\cite{Riess:2019cxk}.
While several observations with different observables from
late time cosmology give consistently a value of about $H_0^{(\text{late})}\approx 74 \km \s^{-1} /\Mpc$,
the expected value extracted from observables dominated by the early universe is about $H_0^{(\text{early})} \approx 67\km \s^{-1} /\Mpc$. This notorious difference
is hard to be understood within the cosmological standard model
or with underestimated observational and systematic uncertainties~\cite{Verde:2019ivm}.
Modifications to the $\LCDM$ model that can resolve the $H_0$-tension are tightly constrained by CMB data \cite{Aylor:2018drw,Knox:2019rjx,Verde:2019ivm}. Nevertheless, several models with early \cite{Poulin:2018cxd,Agrawal:2019lmo}, 
additional self-interacting light relics \cite{Kreisch:2019yzn} among other ideas, such as phantom dark energy relevant at late cosmological time \cite{Alestas:2020mvb}, have been proposed to give an explanation of this discrepancy, see for instance \cite{Knox:2019rjx}.

Let us explain which of the usual assumptions in standard cosmology might be behind the observed $H_0$-tension and why weakening
this assumption might help to release the tension.
While the SM couplings are well understood to be constant at  astronomical scales, this is not necessarily true for gravity 
due to the infrared instability of the graviton propagator~\cite{Houthoff:2017oam,Wetterich:2018qsl,Bosma:2019aiu}, or in other words,
no one really knows how quantum gravity really works.
Thus, one should consider the possibility of
variable cosmological couplings from the Planck-scale~\cite{Bonanno:2017pkg,Reuter:2012xf,Bonanno:2001xi}
down to astronomical scales .

A number of cosmological applications have been considered in the literature. Running vacuum models take advantage of the renormalization group (RG) techniques of quantum field theory (QFT) in curved spacetimes. These models with a decaying vacuum energy density were proposed to tackle the cosmological constant problem in Big-Bang cosmology \cite{Lima:1994gi,Lima:1995ea}. The underlying idea of this formalism is related to earlier attempts addressing the cosmological constant problem, see \cite{Weinberg:1988cp,Sahni:1999gb,Padmanabhan:2002ji,Peebles:2002gy,Sola:2013gha} and references therein.

Another, similar approach, usually applied to black hole physics, but recently used in cosmological models and relativistic stars is scale--dependent gravity 
\cite{Koch:2010nn,Contreras:2013hua,Koch:2013rwa,
Koch:2014joa,Koch:2015nva,Koch:2016uso,Rincon:2017ypd,Rincon:2017goj,
Rincon:2017ayr,Contreras:2017eza,Rincon:2018sgd,Hernandez-Arboleda:2018qdo,
Contreras:2018dhs,Rincon:2018lyd,Rincon:2018dsq,Contreras:2018gct,Canales:2018tbn,Rincon:2019cix,Rincon:2019zxk,Contreras:2019fwu,Fathi:2019jid,Rincon:2019ptp,Contreras:2019cmf,Panotopoulos:2020zqa,Panotopoulos:2021tkk,Panotopoulos:2021obe,Rincon:2021hjj}. 
This approach captures the main idea of
all effective field theory approaches to quantum gravity,
such as the functional renormalization group approach, 
by generalizing the constant couplings of the classical
gravitational action to running, and thus scale--dependent, couplings.
 This method offers a generalization of the Einstein field 
 equations where classical General Relativity (GR) can 
 be recovered in the limit of constant gravitational couplings.
However, the equations of motion obtained from scale--dependence
alone are incomplete. 
In order to complete this set of equations one either needs
input from quantum-gravity calculations (e.g. the beta functions),
or one imposes physically reasonable auxiliary condition(s).

In many applications of GR 
such conditions have already been developed.
The most well-known cases are the energy conditions (for a review see~\cite{Kontou:2020bta}).
One can distinguish between different cases of pointwise energy conditions known as the strong energy condition, the weak energy condition, the dominant energy condition, and the null energy condition. 
The role of these conditions is twofold. On the one hand, they offer a very powerful tool to bypass complicated calculations, which allows deriving general statements such as the famous singularity theorems~\cite{Penrose:1964wq,Hawking:1969sw}.
Supporting this approach, it was shown that some energy conditions, such as the null energy condition, can be rigorously derived in certain theories~\cite{Wall:2009wi,Parikh:2014mja}.
On the other hand, energy conditions have been questioned
as over-simplifications, which can not be expected to hold for all classical and quantum contributions to the stress-energy tensor~\cite{Epstein:1965zza,Visser:1999de,Barcelo:2002bv,Arefeva:2006ido,Curiel:2014zba}.
This triggered an ongoing discussion on the validity of the pointwise energy conditions
and of  improved versions such as 
the quantum energy conditions (see e.g.~\cite{Ford:1978qya,Fewster:2010gm,Ecker:2017jdw,Grumiller:2019xna,Ecker:2019ocp}).
In particular, the stress-energy tensors of different types of matter can be expected to obey different energy conditions. It is therefore quite natural to expect that 
certain energy conditions only hold for certain contributions of -
and not for the total (summed) energy tensor.
In this paper, we will recur to
a particular version of the null energy condition,
which is the least restrictive of the pointwise energy conditions,
in the context of scale-dependent cosmology~\cite{Rincon:2017ayr}. This condition
will be applied only to the novel contribution ($\Delta t_{\mu \nu}$) to the standard stress-energy tensor.
This allows exploiting the full simplifying power of this condition, at the cost of losing some generality
coming from systems that do not fall into this class of theories.

With the scale--dependent formalism plus the application of the null energy condition, the system can be solved (as we will show below).  In the present paper we will study a scale--dependent cosmological model that is not an ad-hoc model invoked to resolve the $H_0$-tension but, as it turns out, the model has appropriate features for facing this problem.

The plan of our paper is as follows. In section \ref{LCDM}, we will review the underlying assumptions and fundamental equations of the $\LCDM$ cosmology. In section \ref{SD}, we will introduce the dynamical equations of the scale--dependent cosmology. A novel
criterion to compare the classical and the scale--dependent Hubble parameter $H(z)$ proposed.
In section \ref{survey} this criterion is then applied 
in a numerical survey of the phase space of the scale--dependent dynamical system. In section \ref{conlusion} we give concluding remarks section and a brief outlook.

For some analytical calculations we have used the
``RGtensors'' code \cite{Bonanos}.

\section{$\LCDM$ cosmology} \label{LCDM}

This section is devoted to review the main features of the $\LCDM$ cosmological model for later reference. The model is based on the assumption of a homogeneous and isotropic universe described by the metric
\begin{align}\label{lineEl}
ds^2 &= -dt^2 + a(t)^2 d\Omega^2\,,
\end{align}
where $a(t)$ is the scale factor, and $d\Omega^2$ is a spatially flat three dimensional space.
The Einstein field equations are
\begin{equation}
 G_{\mu\nu}-\Lambda g_{\mu\nu}=\kappa T_{\mu\nu}\,,
\end{equation}
where $\kappa = 8\pi G$ and $\Lambda$ is the cosmological constant and the contents of the Universe are described by a perfect fluids,
\begin{align}
    T^{\mu \nu} &\equiv (p + \rho)u^{\mu}u^{\nu} - p g^{\mu \nu}\,.
\end{align}

\begin{figure}[ht]
\centering
\includegraphics[width=0.78\textwidth]{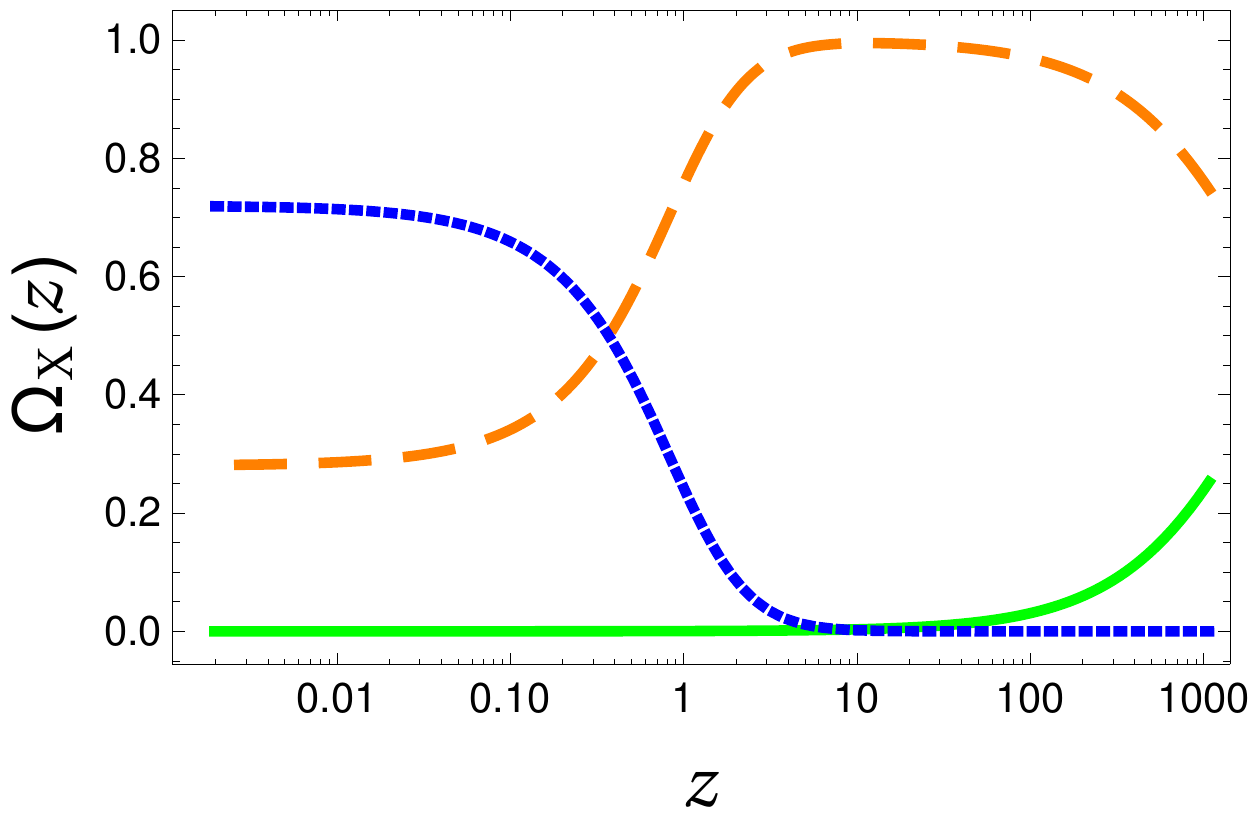}   \
\caption{
Evolution of the density parameters as functions of redshift for the fiducial values (\ref{dps1}), (\ref{dps2}) and (\ref{dps3}). %
The color code is given as follows:
solid green line for radiation $\Omega_r$,
long dashed orange line for the matter term $\Omega_m$
and short dashed blue line for the cosmological term $\Omega_\Lambda$.
}
\label{denparfig}
\end{figure}

For the metric (\ref{lineEl}) these equations reduce to
\begin{align}\label{Feq}
\frac{H^2}{H_0^2} &= \Wl + \frac{\Wr}{a^{4}} + \frac{\Wm}{a^{3}}\,, 
\end{align}
where the Hubble parameter is defined by $H(t) \equiv \dot{a(t)}/a(t)$. In (\ref{Feq}), one uses commonly the density parameters at present: $\Wl$ is the vacuum density, $\Wr$ is the radiation density, and $\Wm$ is the matter density. The density parameters are defined as the ratios of the density to the present critical density
\begin{align}
\Omega_X &\equiv \frac{\rho_{X}}{\rho_{c}}\,, \quad \rho_c=\frac{3H_0^2}{8 \pi G}\,,
\end{align}
where $X = \{\Lambda, r, m \}$ and $H_0$ is the Hubble constant. The Planck collaboration has determined $H_0 = 67.4 \pm 0.5 \km \s^{-1} \Mpc^{-1}$ \cite{Aghanim:2018eyx}, and 
\begin{align} \label{dps1}
\Wm=0.315\pm 0.007\,.
\end{align}
We will also use $\Omega_\gamma =2.47 \times 10^{-5} h^{-2}$ \cite{Fixsen:2009ug} with $h=0.674$ which give us 
\begin{equation}\label{dps2}
\Wr=9.14 \times 10^{-5}\,, 
\end{equation}
if we assume exactly three effective families of neutrinos. The data is compatible with a spatially flat universe which implies
\begin{equation}\label{dps3}
 \Wl = 1-\Wm-\Wr\,.
\end{equation}
The density parameters (\ref{dps1}), (\ref{dps2}) and (\ref{dps3}) are going to be our fiducial values for the rest of the paper. The Friedmann equation (\ref{Feq}) governs the evolution of the scale factor provided the initial condition on $H_0$ and the values for the density parameters, see figure \ref{denparfig}.

%

\section{Cosmological equations in the scale--dependent scenario} \label{SD}

\subsection{General considerations}

This section summarizes the main features of the scale--dependent gravity formalism that allow to derive cosmological equations of this approach. 
The scale--dependent gravity theory applied in this paper 
has been explored 
in a collection of papers mainly focused on black hole solutions 
(see for instance \cite{Rincon:2017ypd,Rincon:2017goj,Rincon:2017ayr,Contreras:2017eza,Contreras:2018dhs}).
Further, certain cases of scale--dependent cosmology have been explored in \cite{Koch:2010nn,Contreras:2013hua,Koch:2013rwa,Koch:2014joa,Koch:2015nva}.
 
The scale--dependent scenario takes the lessons learned from quantum theories in the sense that the coupling constants of the action cannot be treated as constants anymore. 
Instead, these quantities become dependent 
on the renormalization scale $k$.
In the context of a gravitational theory this
means that
\begin{align}
    \{ G_0, \Lambda_0\} \rightarrow \{ G_k, \Lambda_k \}.
\end{align}
Here, $G_0$ stands symbolic for the fixed coupling constants of the classical theory, while $G_k$
stands for the couplings of the scale-dependent theory. From a renormalization group perspective one expects that the scale-dependent coupling $G_k$
approaches the classical coupling $G_0$ in the infrared limit.
At the level of the well known effective action
approach to quantum field theory,
observables and quantum background solutions can be derived from
\begin{equation}\label{actk}
\Gamma[g_{\mu \nu},k] = \int \mathrm{d}^4 x \sqrt{-g} \Bigg[ \frac{1}{2 \kappa_k} \Bigl(R-2\Lambda_k \Bigl) \ + \ \mathcal{L} \Bigg],
\end{equation}  
where $\kappa_k \equiv 8 \pi G_k$ is the Einstein coupling, $G_k$ the Newton coupling, $R$ is the Ricci scalar, $\Lambda_k$ is the cosmological coupling and $\mathcal{L}[g_{\mu\nu},\Psi]$ is the matter Lagrangian, which is taken to have constant couplings at cosmological distance scales. Here, $\Psi$ represents collectively represents matter fields. 
In this formalism, it is essential for
a meaningful physical prediction
to relate the  coarse-graining scale $k$, with 
some physical variable such as
the time variable in cosmology.
This must be taken into account when 
varying (\ref{actk}) with respect to~$\delta g_{\mu \nu}$. The corresponding
equations of motion then read~\cite{Reuter:2003ca,Koch:2010nn,Domazet:2012tw,Koch:2014joa,Contreras:2016mdt}
\begin{equation}
\label{generalized einstein equations}
G_{\mu \nu} + \Lambda_k g_{\mu \nu} = \kappa_k T_{\mu \nu}^{\text{effec}},
\end{equation}
where the corresponding effective energy--momentum tensor is defined as follows
\begin{align}\label{Tefec}
\kappa_k T_{\mu \nu}^{\text{effec}} &\equiv \kappa_k T_{\mu \nu} - \Delta t_{\mu \nu}.
\end{align}
Here, $T_{\mu \nu}$ is the stress-energy tensor of matter, which needs to be conserved.
The tensor $\Delta t_{\mu\nu}$ encodes the scale--dependence of the gravitational coupling.
It is given by 
\begin{align} \label{deltatmunu}
\Delta t_{\mu\nu} &= G_k \Bigl(g_{\mu \nu} \square - \nabla_{\mu} \nabla_{\nu}
\Bigl)G_k^{-1}.
\end{align}
The latter expression clarifies how $\Delta t_{\mu \nu}$ is related to the running of $G_k$ and also implies that when $G_k$ goes to $G_0$ the classical background is unperturbed.

As a consequence of Bianchi's identities, the Einstein tensor's divergence vanishes, $\nabla^{\mu}G_{\mu\nu}=0$. The diffeomorphism invariance can be seen explicitly at the level of the effective action~(\ref{actk}), which only contains invariant quantities. Furthermore, the action associated with the matter sector is assumed to be invariant under general coordinates transformations. Since (i) the matter Lagrangian in (\ref{actk}) is a scalar quantity minimally coupled to the metric field and does not depend on the renormalization scale $k$, and (ii) the $\Psi$-dependence is contained in $\mathcal{L}$ only, the energy-momentum tensor conservation follows from the invariance of $\mathcal{L}$ under general coordinates transformations, as explained in~\cite{Chauvineau:2015cha}. Unless $\mathcal{L}$ contains all the scalar contributions to the action, the independence of $\mathcal{L}$ with the renormalization scale is necessary to retain the conservation equation for the stress-energy tensor (for a detailed discussion, see Appendix A of~\cite{Chauvineau:2015cha}). As a result, $T_{\mu\nu}$ is conserved at the classical level, implying $\nabla^{\mu}T_{\mu\nu}=0$. Thus, the matter energy-momentum conservation implies the canonical scaling of the energy density of matter ($\rho_{m} \propto a^{-3}$) and radiation ($\rho_{r}\propto a^{-4}$), which will be used in the modified Friedman equations (see~(\ref{SD1}),(\ref{SD2})). Since the 
Einstein tensor is conserved $\nabla^{\mu}G_{\mu\nu}=0$, the additional requirement $\nabla^{\mu}T_{\mu\nu}=0$ leads to 
consistency condition that must be met by the metric
$G$ and $\Lambda$. For the line element (\ref{lineEl}) this condition reads
\begin{align} \label{consistencycondition}
\dot{\Lambda}-8\pi T^{00} \dot{G} + \frac{3\dot{G}}{aG^2}\left(\dot{a}\dot{G}+\Ddot{a}G\right)=0.
\end{align}
This relation can be verified for the numerical solutions obtained below.

Equation (\ref{generalized einstein equations}) is 
the generic gap equation of all
effective actions that have the form of~(\ref{actk}).
It is important to note that these equations are incomplete
until an additional equation is imposed.
This additional equation could come from
specifying 
the beta functions of the quantum theory 
and from imposing a scale--setting condition $k\rightarrow k(x)$~\cite{Koch:2010nn,Koch:2014joa}.
Unfortunately, such a procedure comes with large theoretical
uncertainties. Nevertheless, for
a homogeneous cosmological background,
the outcome
of such a procedure will make the
cosmological coupling and the gravitational coupling time-variable functions $G=G(t)$
and $\Lambda=\Lambda(t)$.\\
In this study we will follow the philosophy of
complementing gravitational equations with energy conditions. 
Thus, we will take $G(t)$ and $\Lambda(t)$
as unknown functions and close the system~(\ref{generalized einstein equations})
by invoking an exact energy condition.
In our case this will be an exact null--energy
condition for the stress--energy tensor 
induced by scale-dependence~\cite{Canales:2018tbn}
\be\label{nec}
\Delta t_{\mu \nu}l^\mu l^\nu=0\,,
\ee
where $l^\mu$ is an arbitrary null vector
in the cosmological background metric.
The idea behind this condition is to 
make the energy--momentum tensor introduced
due to scale-dependence, as vacuum-like as possible~\cite{Canales:2018tbn}. 
Typically, energy conditions are used
in terms of inequalities. 
Implementing such an inequality condition into (\ref{nec})
will be discussed after introducing
the explicit equations of motion of the system. \\
The imposed energy condition (\ref{nec}) closes
the system (\ref{generalized einstein equations}) and allows for a straight forward
and predictive implementation of scale-dependence in cosmology.

\subsection{Background evolution}\label{backgroundevolution}

The system of differential equations 
which determines the cosmological background evolution 
is formed by three equations, two Friedmann equations from~(\ref{generalized einstein equations}) and one auxiliary ordinary differential equation obtained from~(\ref{nec}):
\begin{align}
&\frac{1}{H_0^2}\left(H^2 - H \frac{\dot g}{g}\right)=\Wl \lambda(t) + \frac{\Wr}{a^{4}} g + \frac{\Wm}{a^{3}} g \,,\label{SD1}\\
&\frac{1}{H_0^2}\left(2  \dot H + 3H^2-2H\frac{\dot g}{g}+2\frac{\dot g^2}{g^2} - \frac{\ddot g}{g} \right)=3 \Wl \lambda(t) - \frac{\Wr}{a^{4}} g\,,\label{SD2}\\ 
& \frac{\ddot g}{g} -H\frac{\dot{g}}{g} - 2 \frac{\dot g^2}{g^2}=0\,.\label{NECCosm}
\end{align}
where
\begin{equation}\label{dimless}
 g(t)=\frac{G(t)}{G_0}\,, \quad \lambda(t)=\frac{\Lambda(t)}{\Lambda_0}\,.
\end{equation}
As in the $\LCDM$ model, the density parameters $\Wm$, $\Wr$, and $\Wl$ describe the contents of the universe. The time dependence of the $\Lambda$ coupling that is implied in the scale--dependent scenario results in time-dependent dark energy
\begin{equation}
 \Wl \lambda(t)=\frac{\Lambda(t)}{3H_0^2}\,.
\end{equation}
The set of differential equations is enlarged with respect to the $\LCDM$ case and therefore initial conditions on $a(t_0)$, $\dot{a}(t_0)$, $g(t_0)$ and $\dot{g}(t_0)$ are required. The function $\lambda(t)$ can be solved algebraically from either (\ref{SD1}) or (\ref{SD2}) and we will keep the closure condition of the density parameters, eq. (\ref{dps3}).

\begin{figure}[ht!]
\begin{center}
\includegraphics[width=0.84\columnwidth]{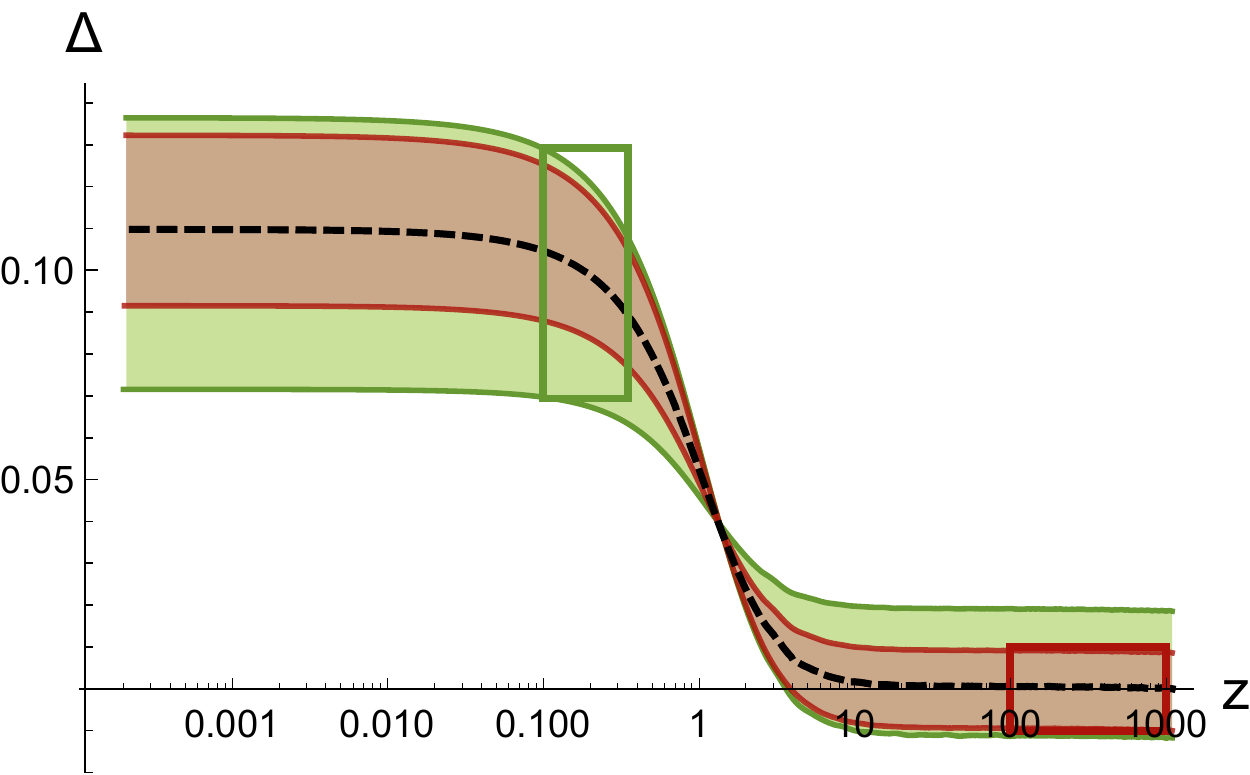} 
\end{center}
\caption{Deviation (\ref{dev}) as a function of $z$. 
The green square indicates
the value and uncertainties associated
with late measurements of the Hubble constant. The brown square indicates
the value extracted from early observables. The brown and green
contours correspond to numerical solutions of (\ref{SD1}-\ref{NECCosm}) with different initial conditions.
The green contours are produced to match
the green square, while the brown contours are produced chosen to match
the red square.}
\label{deltaz}
\end{figure}

Benchmarking the initial data
\begin{align}
&a(t_0)=1\,, \quad \dot a(t_0)=H_0\,,\label{idata1}\\
&g(t_0) = 1\,, \quad\dot g(t_0) \sim 0.1\,,\label{idata2} 
\end{align}
for the fiducial values (\ref{dps1}), (\ref{dps2}) and (\ref{dps3}) gives a noticeable promising solution, see figure \ref{deltaz}. We find that the Hubble parameter $H(t)$ is larger than the Hubble parameter implied by $\LCDM$ at low redshifts, while the density parameters as functions of redshift remain very close to the $\LCDM$ behavior implied by (\ref{Feq}) shown in figure \ref{denparfig}.

The potential of the scale--dependent cosmology for releasing the tension between early--time and late--time measurements of $H_0$ can be assessed by the following function of the redshift implied by the $\LCDM$ model 
\be\label{dev}
\Delta(z) \equiv \frac{H(z)}{H^{(\LCDM)}(z)}-1\,,
\ee
where $H(z)$ is to be evaluated using the scale--dependent cosmology and $H^{(\LCDM)}(z)$ is to be evaluated using (\ref{Feq}). 
For example, the conflicting measurements of the Hubble constant $H_0$ give
\be\label{deltens}
\Delta=\frac{H_0^{(\text{late})}}{H_0^{(\text{early})}}-1 \approx 0.09\pm 0.02\,.
\ee

For given initial conditions, such as (\ref{idata1}) and (\ref{idata2}),  a scale--dependent model would be a good candidate for releasing the tension if it predicts $\Delta \approx 0$ prior to recombination (for the sake of concreteness, let's say for $z\sim 10^3$) and value consistent with (\ref{deltens}) for small $z$.
Figure \ref{deltaz} shows $\Delta$ as a function of $z$ for some
exemplary scale--dependent evolution scenarios.

The figure~(\ref{deltaz}) shows the numerical evaluation of $\Delta(z)$ for initial conditions (\ref{idata1}) and (\ref{idata2}). Green and the brown contours overlap largely, meaning that (\ref{deltens}) is a natural outcome of the discussed SD scenario.

\begin{figure}[ht!]
\begin{center}
\includegraphics[width=0.8\columnwidth]{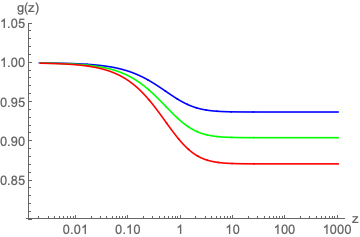} 
\end{center}
\caption{ Blue: Dimensionless gravitational coupling as a function of the redshift (\ref{fig:gz}) for the initial conditions $\dot g(0)=0.06,\;0.11,\; 0.15$.}
\label{fig:gz}
\end{figure}

For the sake of argument in section \ref{perturbations}, let us comment on the behavior of the dimensionless gravitational coupling $g(z)$ as a function of the redshift. Fiducial values for the density parameters and the simplified initial conditions (\ref{idata1}-\ref{idata2}) produce the results shown in figure \ref{fig:gz}. One notes that, for large redshift, the departure from the constant behavior is negligible. This is a direct consequence of the energy condition (\ref{NECCosm}) that gives us
\be\label{gsolpart}
g(t)=\frac{C_1}{C_2+\int^t_{\overline{t}} dt' a(t')}\,.
\ee
From this equation, we see that $g(t)$ is nearly constant when the scale factor is sufficiently small, and $t$ is sufficiently close to a reference time $\overline{t}$. This also tells us that the effects from scale--dependence are expected to be dominant in the late evolution--time. The qualitative behavior as a function of redshift displayed in figure \ref{fig:gz} represents all other initial data explored below. One notes that relative changes of $g(z)$ are small and that $g(z)$ becomes constant for redshifts $z\gtrsim 10$.

\subsection{Energy condition as inequality}\label{sec:encond}

It is natural to generalize the energy condition
(\ref{nec}) to an inequality
by adding an additional parameter (or term $f=f(t)$) to the right-hand side, giving
\be\label{necmod}
\Delta t_{\mu \nu}l^\mu l^\nu=f\,.
\ee
Here, the sign of $f$ determines in which direction the
inequality goes.
The standard null energy condition (NEC) can be cast in the inequality
\be\label{fle0}
f|_{NEC, gen}\le 0.
\ee

The modified NEC (\ref{necmod}) changes the auxiliary condition
for the cosmological framework (\ref{NECCosm}) to
\be\label{NECCosmmod}
\frac{\ddot g}{g} -H\frac{\dot{g}}{g} - 2 \frac{\dot g^2}{g^2}=f.
\ee
For a given evolution of the Hubble parameter,
this relation determines the evolution of the dimensionless
gravitational coupling $g$ (see e.g. (\ref{gsolpart})).
By numerically studying the relation (\ref{NECCosmmod}) 
for a realistic cosmological $H$ evolution, it turned out
that conditions of the type (\ref{fle0}) tend to
produce negative $g$ at  late or  early times of the evolution. The origin of this nonphysical behavior becomes visible, if one considers for example (\ref{NECCosmmod}) at
a starting point $t=t_i$ where $\dot g|_{t_i}=0$. 
At this point, the second derivative  will
be negative $\ddot g<0$, which means that $g(t)$ has a maximum and will run
towards smaller, possibly negative values in the future and/or past.
We can also decrease the probability to fall into nonphysical scenarios with negative $g$ utilizing the following inequality  
\be\label{fge0}
f|_{g, \text{sign}}\ge 0.
\ee
Thus, one has to find a balance between
imposing an inequality in the spirit of energy conditions such as 
(\ref{fle0}) and imposing an inequality in order to
keep the  gravitational coupling positive (\ref{fge0}). 
A potential solution, which accounts for both conditions, is found if one takes advantage of a saturated version of the null energy condition, which means
\be \label{sat_nec}
f \equiv 0
\ee
Such restriction has been consistently used in other works where pure gravity problems are investigated in light of scale-dependent gravity. Thus, we will follow such a route, as it is done in (\ref{NECCosm}) and throughout the rest of the paper.
Although the later condition, i.e., \eqref{sat_nec}, could be highly restrictive, its use offer us a non-trivial property:  
it stabilizes solutions with $\dot g=0$, which is a desirable feature given the strong observational constraints on variable $g$ discussed in section \ref{sec:dG}.

When using this relation one has to keep in mind that neither (\ref{fle0})
nor (\ref{fge0}) are strict inequalities. Like most energy relations, they are reasonable
inequalities. One can typically cook up cases, or a matter contents, where particular energy inequalities (like \ref{fle0}) do not hold exactly, even though they do hold for most other cases. 
Similarly, one can tune particular initial conditions such that negative values of $g$
are avoided, or shifted to the far far future, even though the inequality (\ref{fge0}) does not hold exactly.
Thus, it is fair to say that the choice of the equality (\ref{NECCosm}) in the tension between
 (\ref{fle0}) and (\ref{fge0}) is neither unique nor mandatory but rather a well motivated case study.
 
 Whether or not this case is realized in nature can only be decided by comparing its
 predictions to observations.
This is why the NEC scenario
and its observational implications will be studied throughout the rest of the paper.
However, let us reinforce that it is well-known the null energy condition is always a suitable property in most of pure gravity scenarios.

\subsection{Like Brans Dicke theory, but not quite}

Equations (\ref{SD1}-\ref{NECCosm}) share a resemblance with Brans-Dicke (BD) gravity with a cosmological constant that has also been used to alleviate the $H_{0}$-tension~\cite{Sola:2019jek,Sola:2020lba,Joudaki:2020shz}. As the main similarity, both models promote the classically constant couplings to variable functions. The gravitational scalar fields in both theories carry energy and momentum and contribute to spacetime curvature. In both scenarios, the field equations reduce to the Friedmann equations for constant gravitational couplings. These similarities imply that the terms of the modified Friedmann equations (\ref{SD1}-\ref{NECCosm}) show the same structure as the BD-field equations.
However, both approaches also present differences.
In BD theory, the couplings are dynamical functions of the cosmic time, while in SD scenarios, the cosmic time arises from the renormalization group improving the EH action
and a subsequent scale setting.
The approach used here is inspired by the renormalization group approach, but it 
is simplified by the use of an energy condition.
Further, this energy condition, 
does not come from a variational principle, and therefore the BD pressure equation does not appear in our formalism.

In this context, let us note that~\cite{Reuter:2003ca} proposes an
 alternative way of matching renormalization group results to a BD approach.

\subsection{Classical Limits}

From all known solutions of the system formed by SD and NEC (\ref{generalized einstein equations}, \ref{nec}) we know that 
the corresponding classical solutions of Einstein's equations are obtained in a particular limit
of the integration constants of the generalized system. 

\subsubsection{Zeroth order in SD}

It is interesting to explore the 
classical limit at the level of the equations~(\ref{generalized einstein equations}, \ref{nec}). This analysis applies independently of whether one deals with 
spherical symmetry, cosmological evolution, or the Newtonian limit\dots \\ In order to see how
the classical Einstein equations are incorporated
into the SD system one can factorize
the SD couplings into a constant and a SD part
\bea
\Lambda(x)&=& \Lambda_0 (1+\epsilon \lambda_1(x))\\
G(x)&=& G_0 (1+ \epsilon g_1(x)).
\eea
Here, $\epsilon$ indicates that we take $\lambda_1$
and $g_1$ as small perturbation parameters.
Now, one can study the system for small effects of SD by separating the two parts. The generalized 
field equations~(\ref{generalized einstein equations}) are then
\be\label{SDeomx}
G_{\mu \nu}+ \Lambda_0 g_{\mu \nu}- 8 \pi G_0 T_{\mu \nu}= G_0 g_1(x) T_{\mu \nu} - \Lambda_0 \lambda_1(x) g_{\mu \nu}- \Delta t_{\mu \nu},
\ee
where
\be\label{Deltatx}
\Delta t_{\mu \nu}= (\Box g_{\mu \nu}- \nabla_\mu \nabla_\nu)g_1(x).
\ee
Thus, if  the ``classical limit condition'' (CLC)
\be\label{ClCond}
 G_0 g_1(x) T_{\mu \nu} - \Lambda_0 \lambda_1(x) g_{\mu \nu}- \Delta t_{\mu \nu}=0,
\ee
is fulfilled,
then all classical solutions to Einstein's equations
can be recovered.

Now one can demonstrate three general conclusions
in the context of the system (\ref{SDeomx}, \ref{nec}) and the classical limit (\ref{ClCond}).
\begin{itemize}
    \item[a)]  The CLC in combination with the NEC implies $g_1=\lambda_1=0$:\\
    Contracting the CLC (\ref{ClCond}) from both sides with a null vector $l^\mu$ one gets
    \be
    l^\mu\left( G_0 g_1(x) T_{\mu \nu} - \Delta t_{\mu \nu}\right)l^\nu=0.
    \ee
    By applying the NEC to the second term in this expression one gets
    \be   \label{someCond1}
    G_0 g_1(x) \left(l^\mu\left(  T_{\mu \nu}\right)l^\nu\right)=0.
    \ee
    Thus, for any matter contribution, different from the cosmological constant, and assuming $G_0\neq 0$ one finds
    $g_1=0$, which implies also that $\Delta t_{\mu \nu}=0$.
    By inserting this back into the CLC (\ref{ClCond}) one finds $\lambda_1=0$.
    Thus, all SD corrections $\lambda_1, g_1$ vanish.
    \item[b)] If $g_1$ is negligible and matter is conserved, then the CLC
    (\ref{ClCond}) holds:\\
    From (\ref{Deltatx}) one realizes that $g_1=0$
    implies $\Delta t_{\mu \nu}=0$.
    Thus, the full equations (\ref{SDeomx}) become
    \be\label{SDeomx2}
G_{\mu \nu}+ \Lambda_0 g_{\mu \nu}- 8 \pi G_0 T_{\mu \nu}=  - \Lambda_0 \lambda_1(x) g_{\mu \nu}.
\ee
    Now one can use general covariance and operate with a covariant derivative on this relation giving
    \be\label{someCovD}
    8 \pi G_0 \nabla^\mu T_{\mu \nu}= \Lambda_0 \lambda_1 \nabla_\nu\lambda_1.
\ee
    Thus, if the matter stress-energy tensor is conserved ($\nabla^\mu T_{\mu \nu}=0$), then $\Lambda_0\lambda_1=0$, which
    implies that the CLC holds necessarily.
    \item[c)] If $g_1$ and $\Lambda_0$ are negligible, then matter is conserved and the space-time evolution is classical:\\
    If $g_1=0$ and $\Lambda_0$ are negligible,
    as one could expect at astro-physical scales, 
    then the CLC relation (\ref{ClCond}) holds directly. Thus, the space-time evolution is classical.
    This conclusion is important, since it guarantees, for example, the existence of 
    a well-defined Newtonian limit.
\end{itemize}
We conclude this section with two comments, one on an exception to the first item ($a)$) and one on the stability of classical solutions.

Exception:\\
The above discussion in a) allows for a loophole.
Namely, if the matter stress-energy tensor
is proportional to a cosmological term
\be
T_{\mu \nu}= K(x) g_{\mu \nu}.
\ee
In this case, relation (\ref{someCond1}) becomes trivial and one can not conclude that $g_1=0$.
Instead the CLC becomes
\be\label{ClCondSpecial}
 \left(G_0 g_1(x) K(x) - \Lambda_0 \lambda_1(x)\right) g_{\mu \nu}=\Delta t_{\mu \nu},
\ee
which can be fulfilled, even for non-vanishing $g_1$.
Thus, in the absence of matter, if $\Delta t_{\mu \nu} \sim g_{\mu \nu}$,
space-time can evolve classically, even though $G$ is not constant. This is exactly what we found in \cite{Canales:2018tbn}.

Stability:\\
The above discussion has shown that, for a given problem, it is always possible to find the classical subset of equations 
within the SD equations, for example, the Poisson equation in the Newtonian approximation. 
At the level of solutions, this can be achieved by choosing the right integration constants.
It would now be interesting to see
whether these classical solutions $g_{\mu \nu}^{0}$ are stable, meta-stable, or unstable against SD perturbations $q_{\mu \nu}$ of the metric
\be\label{expandg}
g_{\mu \nu}=g_{\mu \nu}^{0}+\epsilon q_{\mu \nu}.
\ee
To answer these questions, one has to implement
the expansion (\ref{expandg})
into (\ref{SDeomx}) and perform
a fully consistent analysis
of this perturbation theory beyond the classical limit.
However,
as it is well known, perturbation theory of gauge-dependent quantities, such as (\ref{expandg}), is a tricky business, which goes beyond the scope of this paper. It is part of our future plans to perform this type of analysis for the transition between classical and SD gravitational theories.

\subsubsection{Newtonian limit}

From from the solar system scale, we can derive constraints on the SD model at first order perturbations. Our solar system can be described approximately by a spherically symmetric metric, centered at the sun,
\begin{equation}
    ds^2=-(1 + 2 \Phi(t,x))dt^2+a^2 (1 - 2\Phi(t,x)) d\vec{x}^2\,,
\end{equation}
where, by symmetry considerations, it is assumed $\Phi(t,x)=\Phi(r)$ and $a=1$. Now we consider first order expansions in a SD parameter $\epsilon$ for $\Phi$, $G$ and $\Lambda$, where $\Phi$, $G$ and $\Lambda$ are treated as functions of the radial coordinate only. At zeroth order in $\epsilon$, $\Phi$ corresponds to the Newton potential, $G$ corresponds to the Newton constant and $\Lambda$ corresponds to the cosmological constant. From the equations (\ref{generalized einstein equations}) and (\ref{nec}) a bound on the perturbations can be obtained.

We will however take an equivalent path, by using the exact spherically symmetric SD solution of (\ref{generalized einstein equations}) and (\ref{nec}), that was found in \cite{Koch:2015nva}. For spherical symmetry, the null energy condition (\ref{nec}) implies the Schwarzschild relation ($g_{00}=-1/g_{rr}$), and
\begin{align}
-g_{00}(r)=&1+3G_0M_0 (\tilde\epsilon/r_{ref}) -\frac{2G_0M_0}{r} -(1 +6 (\tilde\epsilon/r_{ref}) G_0M_0)(\tilde\epsilon/r_{ref}) r \nonumber\\
&-\frac{\Lambda_0 r^{2}}{3}+ r^{2}(\tilde\epsilon/r_{ref})^{2}(6(\tilde\epsilon/r_{ref}) G_0M_0+1 )\ln \left(\frac{c_4((\tilde\epsilon/r_{ref}) r +1)}{r}\right)\,,
\end{align}
where $r_{ref}$ is a reference length scale and $\tilde \epsilon$ is a dimensionless integration constant of the SD model that can be accounted as a free parameter in this context.

Up to first order in the perturbation, the corresponding potential in the Newtonian limit can be identified as
\be\label{PNpot}
\Phi(r)= -\frac{G_0 M_0}{r}+\tilde \epsilon \frac{3 G_0 M_0-r}{2 r_{ref}}\,,
\ee
where $M_0$ is the mass of the sun. Note that $\tilde \epsilon$ can not be identified with the $\epsilon$ of the previous section in a one to one way because of the restricted aplitacion of the present section. However, keeping the same symmetries (e.g. spherically symmetric) $\tilde \epsilon$ can be compared for systems of the same type. This will be done below.

For the case of our solar system it is convenient to choose the reference scale to be the size of the largest planetary orbit $r_{ref}=r_{ss} \approx 100$ au .
Thus, for $r < r_{ss}$ the dominant contribution to deviations from the classical Newtonian potential is
\be\label{PNestimate}
\delta \Phi = -\tilde \epsilon \frac{r}{2 r_{ss}}.
\ee
As a rough back of the envelope estimate, $\delta \Phi$ should be smaller than
the typical dimensionless post-Newtonian parameters (like $\gamma-1$) obtained from our solar system, which are of the order of $10^{-5}$~\cite{Will:2014kxa}. Thus, one finds from (\ref{PNestimate}) that
\be\label{boundSS}
|\tilde \epsilon|_{ss} \times \left( \frac{r}{r_{ss}}\right)\lesssim 10^{-5}.
\ee
Therefore, at solar system scales, $r \lesssim r_{ss}$, the strongest upper bound is $\tilde \epsilon|_{ss} \lesssim 10^{-5}.$

At sub-cosmological distance scales, the linear $r$ term in (\ref{PNpot}) is an indication
that tiny SD effects in our solar system can be expected to become relevant at much larger distances.
 Although interesting, this goes beyond the scope of the present paper and it will be studied in a separate paper.

\subsection{Cosmological perturbations in the SD scenario}\label{perturbations}

The standard approach to cosmology perturbations splits the metric (\ref{lineEl}) in linear perturbations $h_{\mu \nu}$ on top of a $\bar g_{\mu \nu}$ background metric
\be\label{metPert}
g_{\mu \nu}= \bar  g_{\mu \nu}+ \epsilon a^2 h_{\mu \nu}\,,
\ee
where the small expansion parameter is $\epsilon\ll1$. Perturbations of the stress-energy momentum tensor are defined by
\be\label{EnMomPert}
T_{\mu \nu}=  \bar T_{\mu \nu}+\epsilon \Theta_{\mu \nu}\,.
\ee
The usual power counting without scale-dependence is the following:
\begin{itemize}
  \item To order $\epsilon^0$, one obtains the background evolution given by the Friedman equations.
  \item To order $\epsilon^1$, one obtains the leading corrections to the background evolution. These equations  describe the evolution of metric and matter perturbations~\cite{Bardeen:1980kt,Fry:1983cj,Kodama:1985bj}. The background scale factor obtained at order $\epsilon^0$ serves as input  for the perturbations at order~$\epsilon^1$
  \item Higher powers such as $\epsilon^2$ are neglected.
\end{itemize}
In the scale--dependent scenario, we focused on benchmarks where the scale--dependent corrections are small. A similar type of perturbative SD cosmology has already been introduced in~\cite{Toniato:2017wmk}.
These corrections can thus be symbolized by a small parameter $\xi$. Our scenarios have thus two expansion parameters, $\epsilon$ and $\xi$, with $\epsilon<\xi\ll1$. Note that for larger red-shift one even has
$\epsilon\approx \xi\ll1$. The perturbative power counting in the case of scale-dependence will thus be
\begin{itemize}
  \item To order $\xi^1$, one obtains the background evolution given by the modified Friedman equations.
  \item To order $\epsilon^1$, one obtains the leading corrections to the background evolution. These equations are identical to the ones in the $\LCDM$ scenario. The only difference is that the input of the background scale factor contains scale-dependence corrections.
  \item Higher powers, such as $\xi^1 \epsilon^1$ or $\epsilon^2$, are neglected.
\end{itemize}
One can thus conclude that for sufficiently small $\xi$ all corrections to the CMB perturbations are generated due to the modified background evolution, and therefore, for \emph{weak scale-dependence}, it is imposed that
\begin{equation}\label{weakscaledependence}
 \Delta(z) < \mathcal{O} (0.1)\,, \quad \text{and} \quad \delta_g < \mathcal{O}(0.1)\,,
\end{equation}
where
\be
\delta_g=g(z)-1\,.
\ee
However, for \emph{larger scale-dependence} $\xi\approx 1$, one can expect significant corrections at the level of the perturbation equations. This type of analysis of large scale--dependence effects goes beyond the scope of this paper and will be postponed to a future study.

\subsection{Bounds on $\dot{G}(t)$}
\label{sec:dG}

A possible time variability of Newton's constant $G=G(t)$
is the subject of numerous experimental tests. Methods are based on very diverse physics and the resulting upper bounds may differ by several orders of magnitude, for extensive reviews see e.g.~\cite{Uzan:2002vq,Uzan:2010pm}. One can distinguish between two types of constraints:
\begin{enumerate}[(a)]
    \item Local constraints, based on inner galactic or astrophysical observations. The stronger local constraints to date come from lunar ranging experiments~\cite{Williams:2012nc,Hofmann:2018myc}, giving $|\dot G /G|=(7.1 \pm 7.6) \times  10^{-14} \yr ^{-1}$; solar system observations \cite{Fienga:2014bvy,Pitjeva:2013xxa}, giving $|\dot G /G|=(0.55 \pm 0.15) \times  10^{-13} \yr ^{-1}$ and $|\dot G /G|=(6.3\pm 4.3) \times  10^{-14} \yr ^{-1}$; and pulsar timing observations~\cite{Deller:2008jx}, giving $\dot G /G=(-5 \pm 26) \times  10^{-13} \yr ^{-1}$. Other local constraints come from pulsar binaries~\cite{Zhu:2018etc}, giving $\dot G /G=(-0.1 \pm 0.9) \times  10^{-12} \yr ^{-1}$; ephemeris of Mercury~\cite{Genova:2018gti}, giving $|\dot G /G| < 4 \times  10^{-12} \yr ^{-1}$; exoplanetary motion~\cite{Masuda:2016ggi}, giving $|\dot G /G| < 10^{-6} \yr ^{-1}$; surface temperature of millisecond pulsars~\cite{Jofre:2006ug}, giving $|\dot G /G| < 4\times  10^{-12} \yr ^{-1}$; and pulsating white-dwarfs~\cite{Corsico:2013ida}, giving $|\dot G /G| \sim 1.8 \times  10^{-10} \yr ^{-1}$.
    \item Constraints based on cosmological observations. The strongest bounds come from big bang nucleosynthesis~\cite{Copi:2003xd}, giving $|\dot G /G| < 4\times  10^{-13} \yr ^{-1}$. Other cosmological constraints come from helioseismology on cosmological time scales~\cite{Guenther:2009xp}, giving $|\dot G /G| \lesssim 1,6 \times  10^{-12} \yr ^{-1}$ and anisotropies in cosmic microwave background~\cite{Wu:2009zb}, giving $|\dot G /G| < 1.05\times  10^{-12} \yr ^{-1}$. We also have constraints coming from gravitational waves \cite{Yunes:2009bv}, giving $|\dot G /G| \lesssim   10^{-11} \yr ^{-1}$; and also gravitational waves + supernovae \cite{Zhao:2018gwk}, giving $|\dot G /G| \lesssim 3 \times  10^{-12} \yr ^{-1}$. The latter can be used to constraint cosmological parameters in a model independent way.
\end{enumerate}

When using local constraints on $\dot G/G$, one has to remember that in the SD approach, $G=G(x^\mu)$ and therefore, the simplification $G=G(t)$ would not apply when considering the inside of local structures. 
Within such local structures,
even if they are spherically symmetric, one would
have to consider a time and radius dependence of the couplings $G=G(t,r)$.
There are no known solutions to such systems with scale-dependence.
Even if one assumes further that 
the time evolution of the couplings is locally suppressed inside of gravitationally bound structures $G(t,r)\approx G(r)$,
the Newtonian limit is not trivial~\cite{Koch:2010nn,Contreras:2013hua,Koch:2013rwa,
Koch:2014joa,Koch:2015nva,Koch:2016uso,Rincon:2017ypd,Rincon:2017goj,
Rincon:2017ayr,Contreras:2017eza,Rincon:2018sgd,Hernandez-Arboleda:2018qdo,
Contreras:2018dhs,Rincon:2018lyd,Rincon:2018dsq,Contreras:2018gct,Canales:2018tbn,Rincon:2019cix,Rincon:2019zxk,Contreras:2019fwu,Fathi:2019jid,Rincon:2019ptp,Contreras:2019cmf,Panotopoulos:2020zqa}.

This means that the local constraints on $\dot G/G$ can not be directly applied to the above large-scale cosmological model of SD.

Cosmological constraints on $\dot G/G$ are a different story for us. Large scale measurements are a priori within the regime of applicability of the above SD cosmological model. However, when such observables are translated into bounds on $\dot G/G$ it is typically assumed that all other cosmological parameters, in particular $\Lambda=\Lambda_0$, are constant and therefore cosmological constraints require to be redrawn in the context of the SD model (in the SD cosmology $\Lambda=\Lambda(t)\neq \Lambda_0$). Therefore we conclude that the list of constraints in (b) do not apply to the SD model presented in this paper. The derivation of new cosmological bounds based on the SD scenario goes beyond the scope of this paper and will be postponed to future work.

We will, however, adopt a conservative approach by focusing mainly on a region of $\dot{G}/G$ that is compatible with the bounds described in (b),
\begin{equation}\label{constraintsGdotG}
 \log_{10} \left|\frac{\dot{G}/G}{\Gyr ^{-1}} \right|< -2,\dots ,-4\,.
\end{equation}
The r.h.s. of (\ref{constraintsGdotG}) ranges from less to most conservative.

\subsection{The age of the Universe}\label{sec:age}

We can also make the distinction between cosmological and local methods of estimating the age of the universe. First, there is the age deduced from an explicit cosmological model such as $\LCDM$. Limits in this case would have to be re-derived within the SD model. Second, a lower limit on the age of the universe can be deduced from the age of the older astrophysical objects by using models of stellar evolution. In both cases the existing bounds of the age of the universe have to be used with care and caution. Limits based on stellar evolution seem more appropriate for us because they do not rely strongly on a particular cosmological model, however, even here one has to be careful. The reason is, that a fully SD theory would in principle depend on spatial and time-like scales. Thus, in local clusters of matter, the gravitational couplings and time evolution can be expected to evolve slightly different from the same couplings in the outer void. For the present paper, however, we will assume that such effects are small.

\section{Numerical survey}\label{survey}

In this section the conceptual ideas
from the subsections \ref{backgroundevolution}
and \ref{perturbations} will be confronted with the observational bounds
discussed in the subsections~\ref{sec:dG}
and~\ref{sec:age}.

\subsection{Degrees of freedom}

We will explore the phase space of the dynamical system given by eqs. (\ref{SD1}-\ref{NECCosm}). For this discussion it is convenient to use the dimensionless functions (\ref{dimless}) and,
\begin{equation}
 x=\frac{a(t)}{a_0}\,,
\end{equation}
The scale-dependent cosmological equations take the form
\begin{align}
 &\frac{x'(\tau)^2}{x(\tau)^2}-\frac{g'(\tau) x'(\tau)}{g(\tau) x(\tau)}= \Wm g(\tau)x(\tau)^{-3}+\Wr g(\tau)x(\tau)^{-4}+\Wl \lambda(\tau)\,,\label{eq1}\\
 &\frac{x'(\tau)^2}{x(\tau)^2}-2\frac{g'(\tau) x'(\tau)}{g(\tau) x(\tau)}+2\frac{g'(\tau)^2}{g(\tau)^2}+2\frac{x''(\tau)}{x(\tau)}-\frac{g''(\tau)}{g(\tau)}=-\Wr g(\tau) x(\tau)^{-4}+3\Wl \lambda(\tau)\,,\label{eq2}\\
 &\frac{g''(\tau)}{g(\tau)}-\frac{g'(\tau) x'(\tau)}{g(\tau) x(\tau)}-2\frac{g'(\tau)^2}{g(\tau)^2}=0\,.\label{eq3}
\end{align}
The evolution of this dynamical system can be determined by giving the initial conditions $x(\tau_0)$, $x'(\tau_0)$, $g(\tau_0)$ and $g'(\tau_0)$, where we denoted $\tau_0$ as the present value of the evolution coordinate $\tau$. This means an increase from one to four d.o.f. with respect to the $\LCDM$ case, where only $x'(\tau_0)$ is required to provide initial values. In the $\LCDM$ case this degree of freedom is quite often traded for the value of $h$.

\begin{figure}[h]
  \centering
  \includegraphics[width=.4\linewidth]{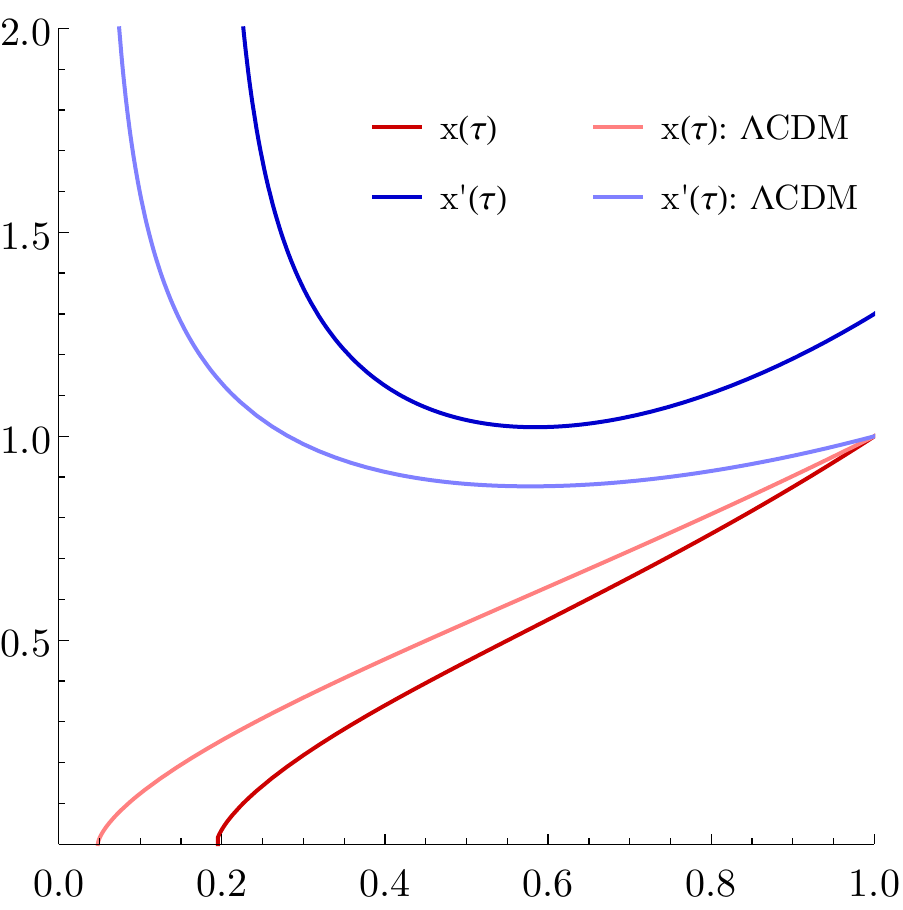}
  \includegraphics[width=.4\linewidth]{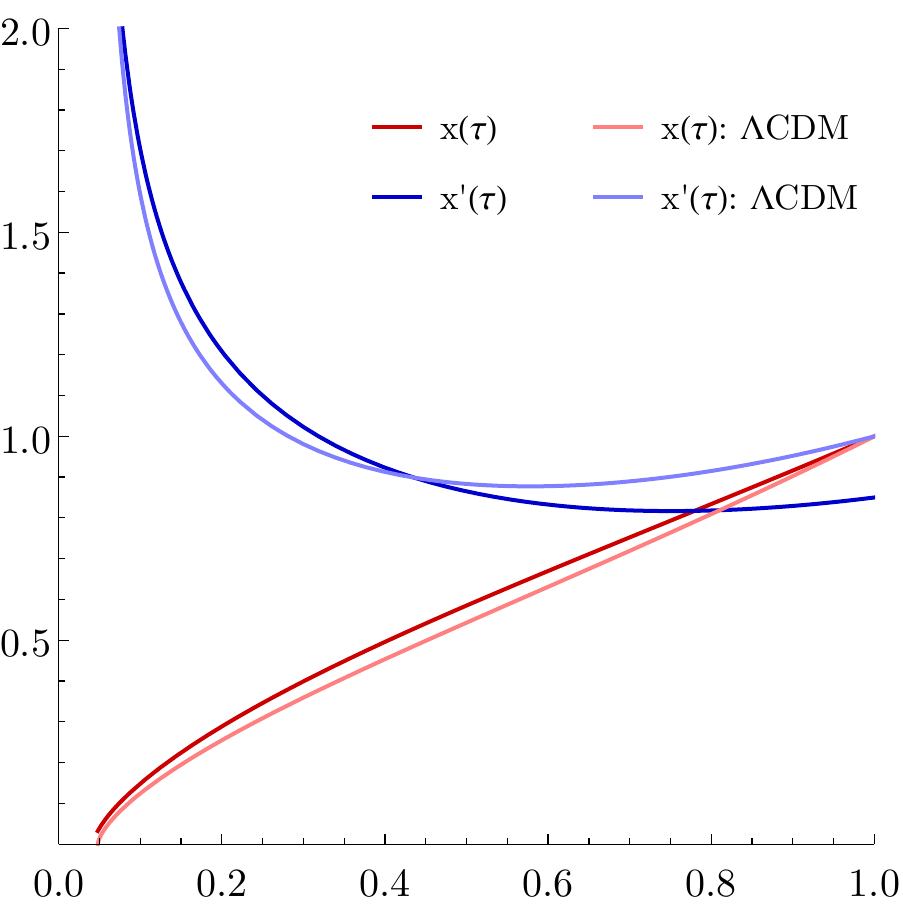}
\caption{Numerical solution of the scale-dependent eqs. (\ref{eq1}-\ref{eq3}) for $g'(t_0)/g(t_0)=-0.5$, $x(t_0)=1$ and $x'(t_0)=1.3 \text{ and } 0.85$ on the left and right plots respectively. For $x'(t_0)/x(t_0)>1$ ($<1$) we have $\tau_\text{age}$ smaller (bigger) than the predicted value from a $\LCDM$ model, however the resulting physical age $t_\text{age}$ (eq. (\ref{tage})) is generically smaller for values of $h$ such that $(x'(t_0)/x(t_0) \ h \ H_{100} )\approx H_0(\text{late})$.}
\label{plotdof}
\end{figure}

On physical grounds, we will set $g(\tau_0)=1$, which implies that we are left with three d.o.f. that are given by $x(\tau_0)$, $x'(\tau_0)$ and $g'(\tau_0)$. These degrees of freedom characterize possible dynamics of the scale--dependent cosmology. For a fixed value of $g'(\tau_0)$, we can visualize $x(\tau_0)$, $x'(\tau_0)$ as the vertical position of the red and blue line at $\tau=\tau_0$ in the plot of Figure \ref{plotdof} . We will trade the $x(\tau_0)$ d.o.f. for the value of $h$, which can be understood as the horizontal displacement of the intersection of the red line with the horizontal axis of the plot. Therefore we will take $x(\tau_0)=1$ and $h$ as a free parameter. The reason for choosing $x(\tau_0)=1$ is that we will use fiducial values from the Planck collaboration for the density parameters $\Wm$ and $\Wr$ to carry out a numerical survey of the phase space of equations (\ref{eq1}-\ref{eq3}). Our basic assumption here is that a complete analysis of CMB anisotropies using the scale--dependent cosmology would determine values for the density parameters that are a small correction from the values inferred from a pure $\LCDM$ model,
as discussed in subsection \ref{perturbations}.

Mathematical consistency of (\ref{eq1}-\ref{eq3}) forbids to take $\lambda(\tau)$ as a constant for generic initial conditions. The function $\lambda(\tau)$, however, does not carry an independent degree of freedom as it can be determined algebraically from any of the eqs. (\ref{eq1}) or (\ref{eq2}) at any giving time. The numerical evaluation of $\lambda(\tau)$ explicitly reveals the time dependence of $\lambda(\tau)$ for generic initial conditions. 

In the next section, we will mostly focus on regions where $g'(\tau)/g(\tau)$ is small enough to avoid conflict with constraints on the time dependence of Newton's constant. Based on the constraints (\ref{constraintsGdotG}) and using $H_{100} \sim (10 \Gyr)^{-1}$, we can express the constraints with respect to the dimensionless time variable $\tau$ and the $h$ parameter,
\begin{equation}
 \log_{10} \left|h g'/g \right|< -1, \dots ,  -4 \,.
\end{equation}

\subsection{Constraints on models}\label{constraintsonmodels}

In this subsection we will put constraints on the values of $\alpha$ and $\gamma$ given by
\begin{equation}\label{alphagamma}
 \alpha= \frac{x'(\tau_0)}{x(\tau_0)}\,, \quad \gamma= \frac{g'(\tau_0)}{g(\tau_0)}\,,
\end{equation}
and $h$ defined as $\tau=h \ H_{100} \ t$ 
by carrying out a numerical survey. 
As benchmark tests we will focus on following criteria:
\begin{enumerate}[(i)]
 \item expansion rate of the Universe at present as implied by late time observations, $H_0^{(\text{late})}$;
 \item expansion rate at $z \sim 10^3$ as it is inferred from CMB anisotropies, $H_0^{(\text{early})}$;
 \item constraints on the age of the Universe, $t^\ast_\text{age}$ coming from stellar evolution.
\end{enumerate}

For criterion (i), we will use measurements of $H_0$ given by late time observations such as SN Ia or Cepheids \cite{Freedman:2000cf}. We are going to use the measurement provided by the SH0ES project that is on the large-side of values, $H_0^{(\text{late})}=74.0\pm 1.4 \km \s^{-1} \Mpc^{-1}$ \cite{Riess:2019cxk}. For illustration purposes, we will use in criterion (ii) the result from the Planck collaboration, $H_0^{(\text{early})}=67.4\pm 0.5\km \s^{-1} \Mpc^{-1}$.  Numerical solutions confirm the ideas in section \ref{backgroundevolution}, and therefore the evaluation of criterion (ii) is not sensitive to the specific choice of redshift as long as it is well inside the matter dominated era, see the figure \ref{deltaz}. For the criterion (iii), we will use an estimate that comes from stellar evolution. Numerical solutions show a general tendency of predicting a slightly smaller $t_\text{age}$ than the $\LCDM$ prediction (see figure \ref{plotdof}), therefore we will take a conservative approach by using an estimate of the age of the Universe that is on the large-side of values $t^\ast_\text{age} = 14.46\pm 0.8 \Gyr$ \cite{Bond:2013jka}.

We will quantify deviations in the predicted values by a given set of initial conditions $(\alpha,\gamma,h)$ with functions $\Sigma_{(i)}$, $\Sigma_{(ii)}$ and $\Sigma_{(iii)}$. The functions $\Sigma$ measure deviations with respect to the central value in units of standard deviations, and to the best knowledge of the authors, these functions are being defined here for the first time. For criterion (i), we used the function
\begin{equation}
 \Sigma_{(i)}=\frac{\text{abs}\left[\left.\frac{\dx/x}{\dx^{(\LCDM)}/x^{(\LCDM)}}\right|_{z\sim0}-\frac{H_0^{(\text{late})}}{H_0^{(\text{early})}}\right]}{\frac{H_0^{(\text{late})}}{H_0^{(\text{early})}}\left(\left(\frac{\Delta H_0^{(\text{late})}}{H_0^{(\text{late})}}\right)^2+\left(\frac{\Delta H_0^{(\text{early})}}{H_0^{(\text{early})}}\right)^2\right)^{1/2}} \,,
\end{equation}
where $x^{(\LCDM)}$ represents the $\LCDM$ prediction evaluated with Planck results. The quantities. The function $\Sigma_{(i)}$ is given relative to $\sigma=1$ uncertainties, $\Delta H_0^{(\text{late})}$ and $\Delta H_0^{(\text{early})}$. If, for example a given model for $(\alpha,\gamma,h)$ predicts,
\begin{equation}
 \left.\dx/x\right|_{z\sim0}=H_0^{(\text{early})}\,,
\end{equation}
then $\Sigma_{(i)}\approx 4.4$, which represents the current tension between early time and late time measurements  and therefore such $(\alpha,\gamma,h)$-model does not have the potential of describing the data any better than the $\LCDM$ prediction. Any model $(\alpha,\gamma,h)$ that is capable of reducing the value of $\Sigma_{(i)}$ would be releasing tension between early time and late time measurements. Models with $\Sigma_{(i)} < 1$ are compatible with both measurements.

The suitability of a given model $(\alpha,\gamma,h)$ for criterion (ii) will be evaluated with the function 
\begin{equation}\label{Sigmaii}
 \Sigma_{(ii)}=\frac{\text{abs}\left[\left.\frac{\dx/x}{\dx^{(\LCDM)}/x^{(\LCDM)}}\right|_{z\sim 10^3}-1\right]}{\Delta H_0^{(\text{early})}/H_0^{(\text{early})}} \,.
\end{equation}
A value $\Sigma_{(ii)} < 1$ ensures compatibility with the results from the Planck collaboration.

For criterion (iii) we will use the function
\begin{equation}\label{Sigmaiii}
 \Sigma_{(iii)}=\frac{\text{abs}\left[t_\text{age}-t^\ast_\text{age}\right]}{\Delta t^\ast_\text{age}} \,.
\end{equation}
A value $\Sigma_{(iii)} < 1$ ensures compatibility age estimates from stellar evolution.

The evaluation of $\Sigma_{(ii)}$ and $\Sigma_{(iii)}$ requires to use a numerical solution for a given set parameters $(\alpha,\gamma,h)$. Thanks to the form of the equations (\ref{eq1}-\ref{eq3}), we can benchmark a set of models $(\alpha,\gamma,h)$ with a single numerical solution of the model $(\alpha,\gamma)$. This is because the relevant quantities in (\ref{Sigmaii}) and (\ref{Sigmaiii}) have well-defined scaling properties,
\begin{align}
 \dx(t)/x(t) =& h H_{100} x'(\tau)/x(\tau)\,,\\
 t_\text{age} =& h^{-1} H^{-1}_{100} \tau_\text{age}\,. \label{tage}
\end{align} 
In (\ref{tage}), $\tau_\text{age}$ is the age of the Universe in the dimensionless time variable $\tau$ that is determined from the numerical solution with the $(\alpha,\gamma)$ initial conditions.

\begin{figure}[h]
  \centering
  \includegraphics[width=.6\linewidth]{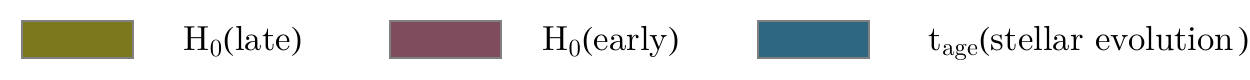}\\
  \includegraphics[trim={10 2 10 8},clip,width=.32\linewidth]{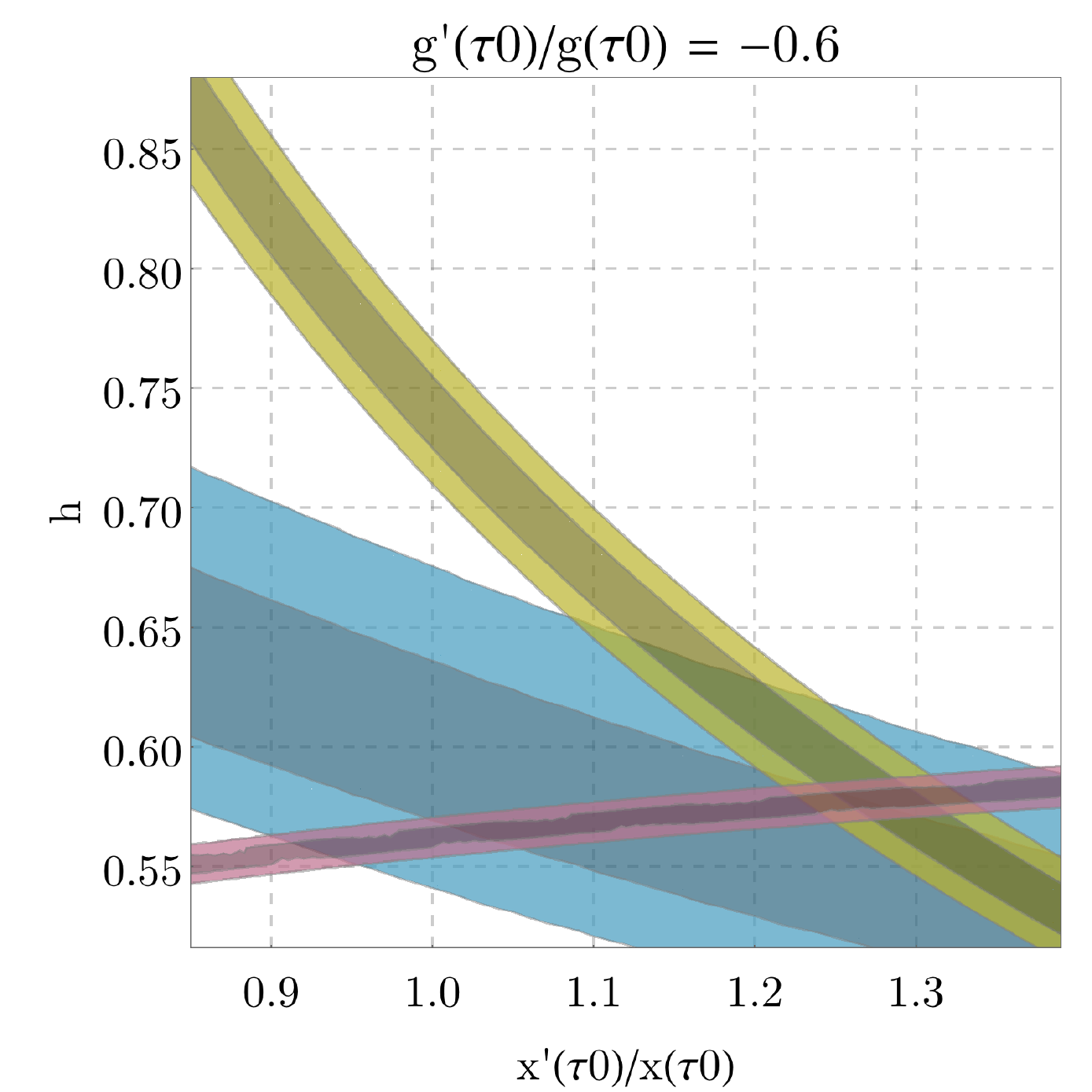}
  \includegraphics[trim={10 2 10 8},clip,width=.32\linewidth]{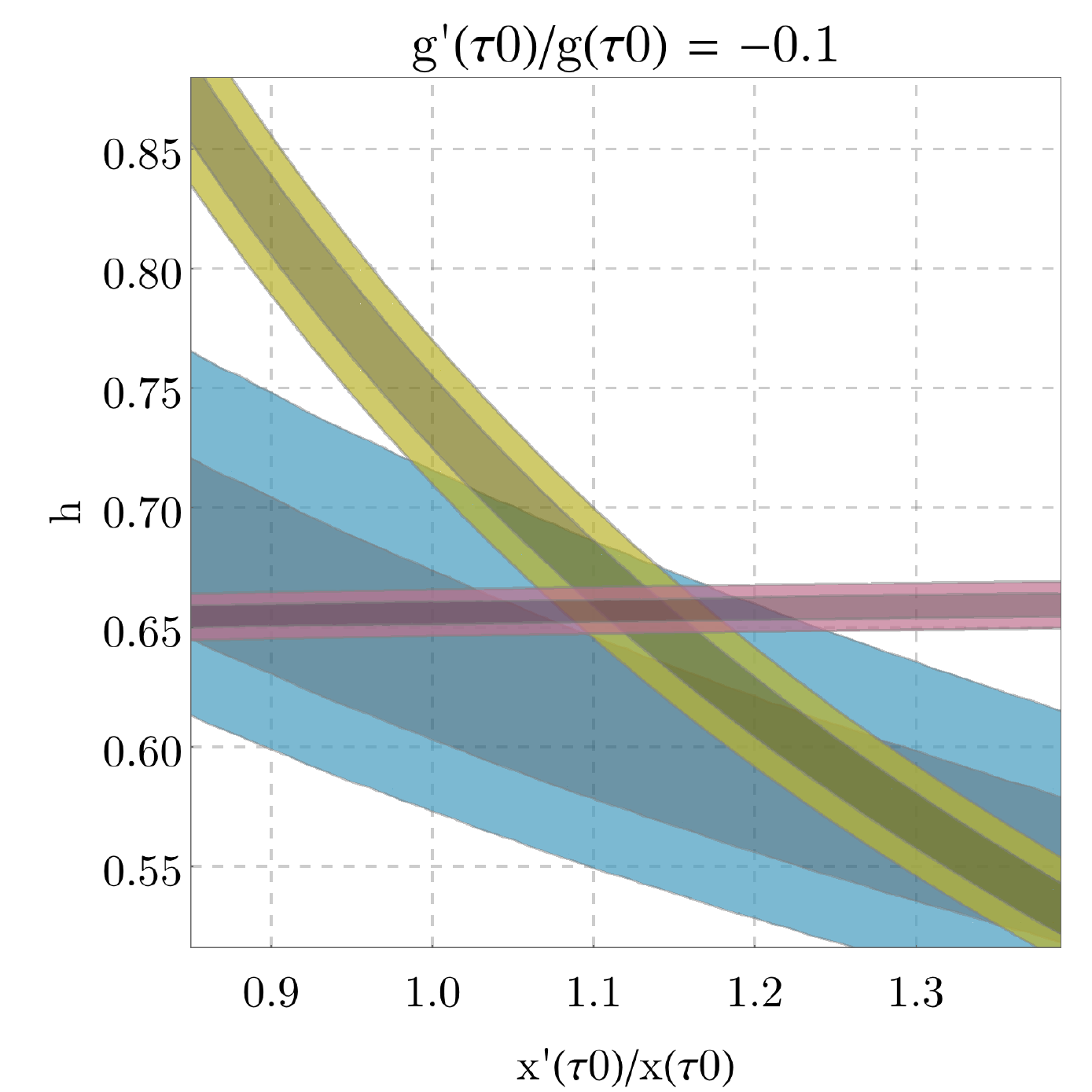}
  \includegraphics[trim={10 2 10 8},clip,width=.32\linewidth]{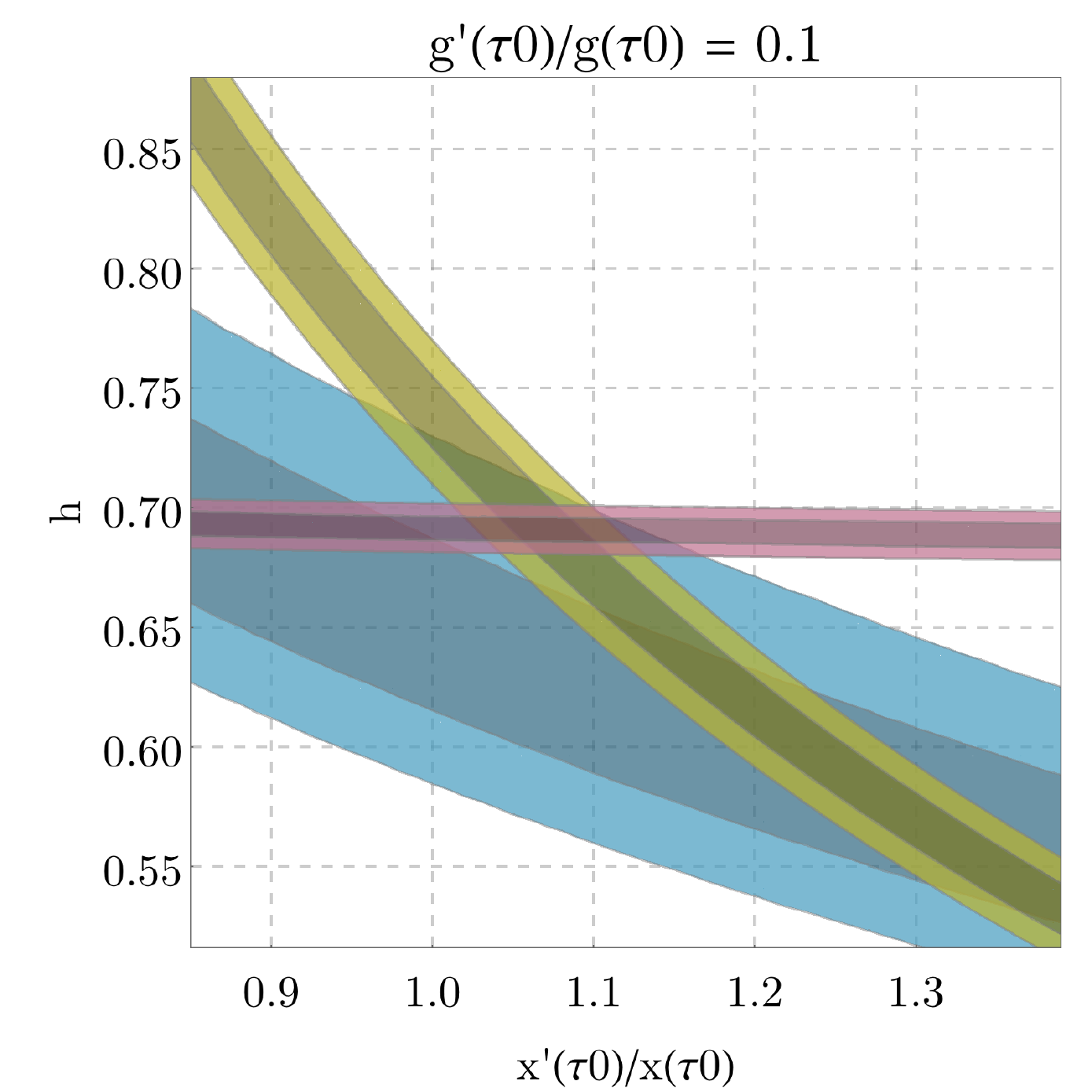}\\
  \includegraphics[trim={10 2 10 8},clip,width=.32\linewidth]{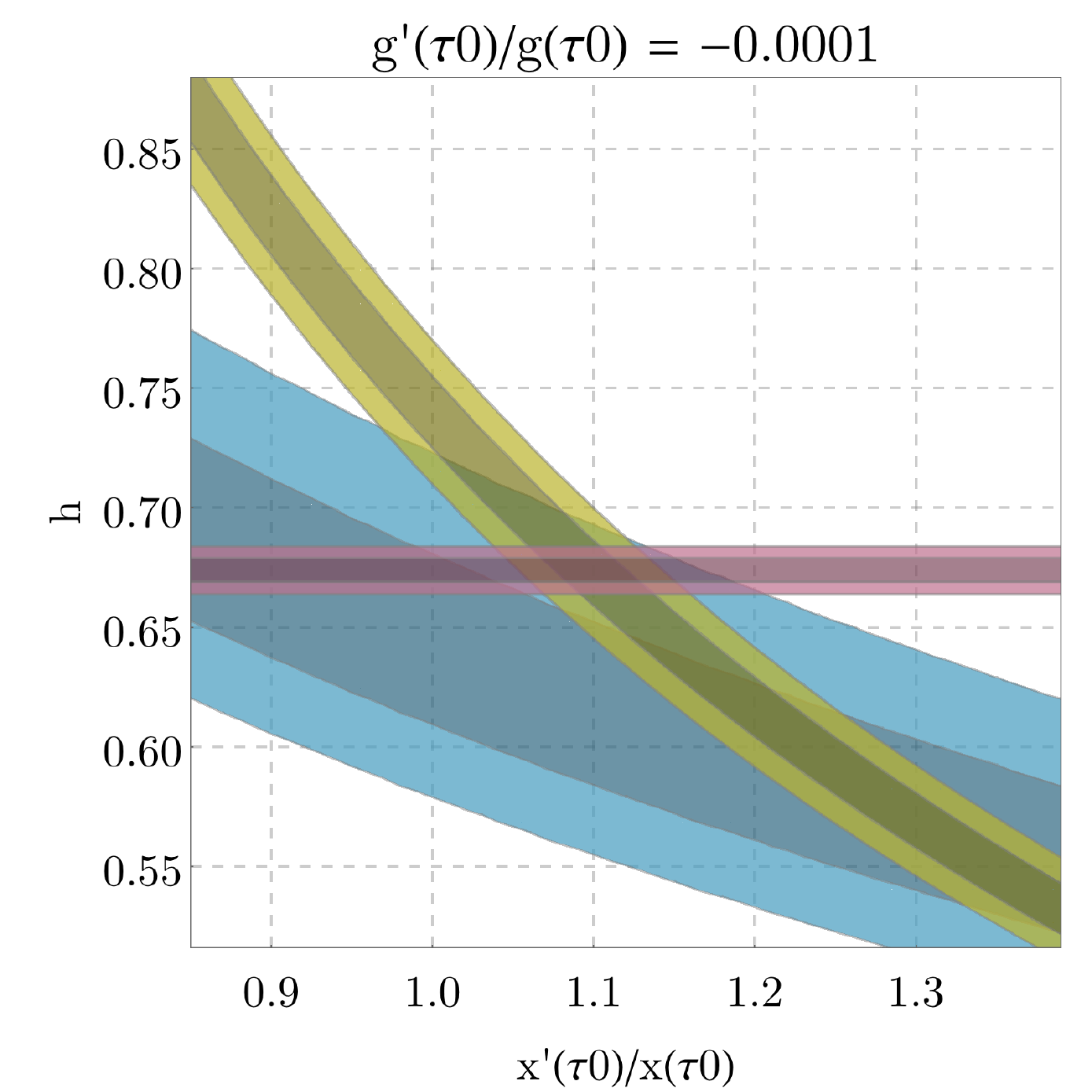}
  \includegraphics[trim={10 2 10 8},clip,width=.32\linewidth]{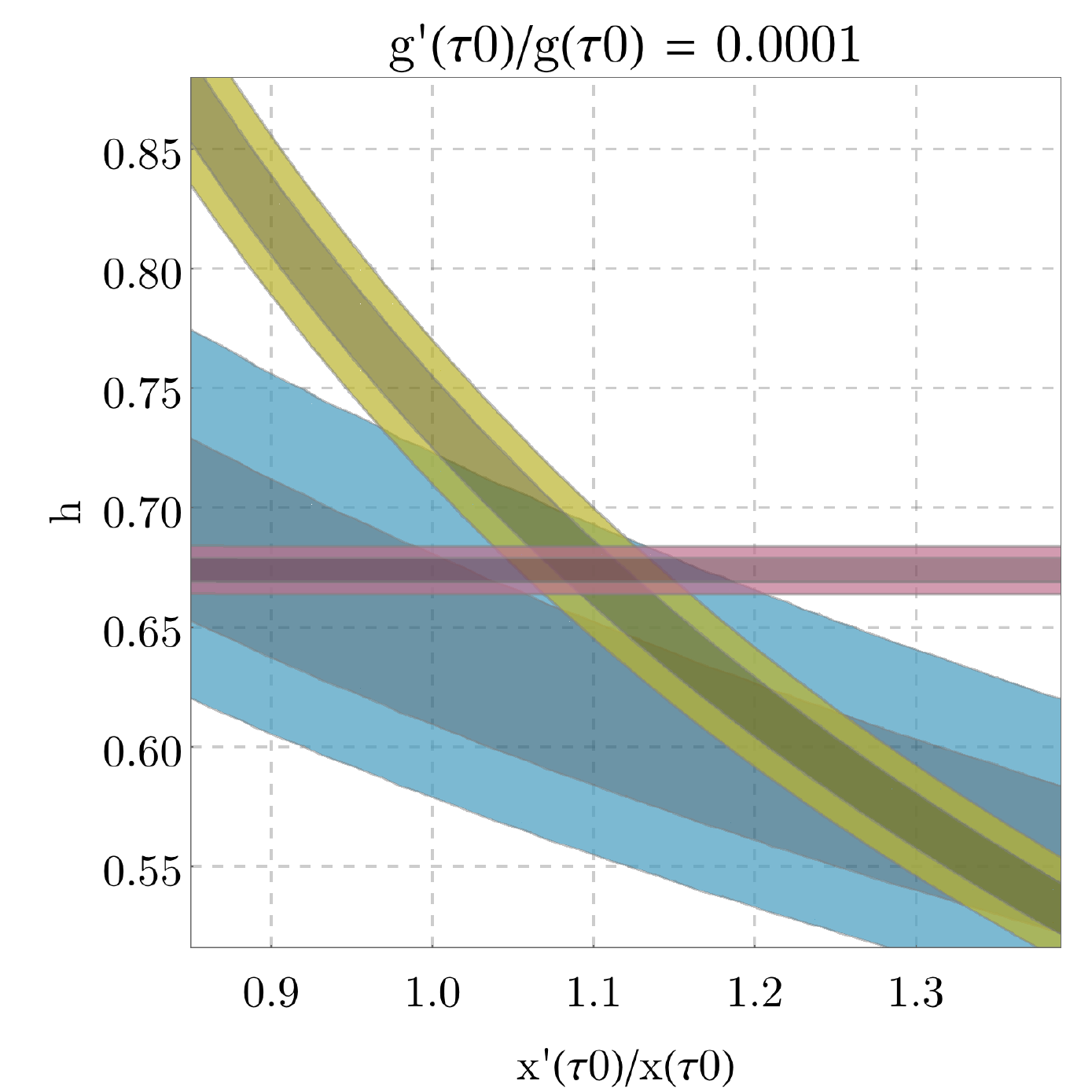}
\caption{Contour plots of $\Sigma < 1$(darker color) and $\Sigma < 2$(lighter color) for fixed values of $\gamma=g'(\tau)/g(\tau)$. Further comments in the text.}
\label{regions1}
\end{figure}

In figure \ref{regions1} we show contour plots of $\Sigma < 1,2$ regions ($\sigma=1,2$ regions resp.) in the $(\alpha,h)$ plane for fixed values of $\gamma$. In the first row we considered values of $h \gamma \lesssim 10^{-1}$ that are in the upper range of constraints \cite{Yunes:2009bv}. This constraints serve as an illustration of the effect of using positive or negative values of $\gamma$. In the second row of figure \ref{regions1} there are plots for $h \gamma \lesssim 10^{-4}$ that respect the strongest constraints in the literature \cite{Williams:2012nc,Hofmann:2018myc}. All plots of figure \ref{regions1} exhibit a triple intersection in the $\Sigma_{(i)} < 1$, $\Sigma_{(ii)} < 1$ and $\Sigma_{(iii)} < 2$ regions, and the case of $\gamma \approx -0.6$ is special for having a small region (but not point-like) of triple intersection $\Sigma_{(i)}, \Sigma_{(ii)}, \Sigma_{(iii)} < 1$. Values $|\gamma| \sim 10^{-4}$ are degenerate with respect to the $\LCDM$ case in the sense that the region of $\Sigma_{(ii)}<1$ contains the point $(\alpha,h)=(1,0.67)$, however the region $\Sigma_{(i)}<1$ does not contain that point, as expected.

\begin{figure}[h]
  \centering
  \includegraphics[width=.6\linewidth]{{plotlegend}.pdf}\\
  \includegraphics[trim={10 1 10 6},clip,width=.32\linewidth]{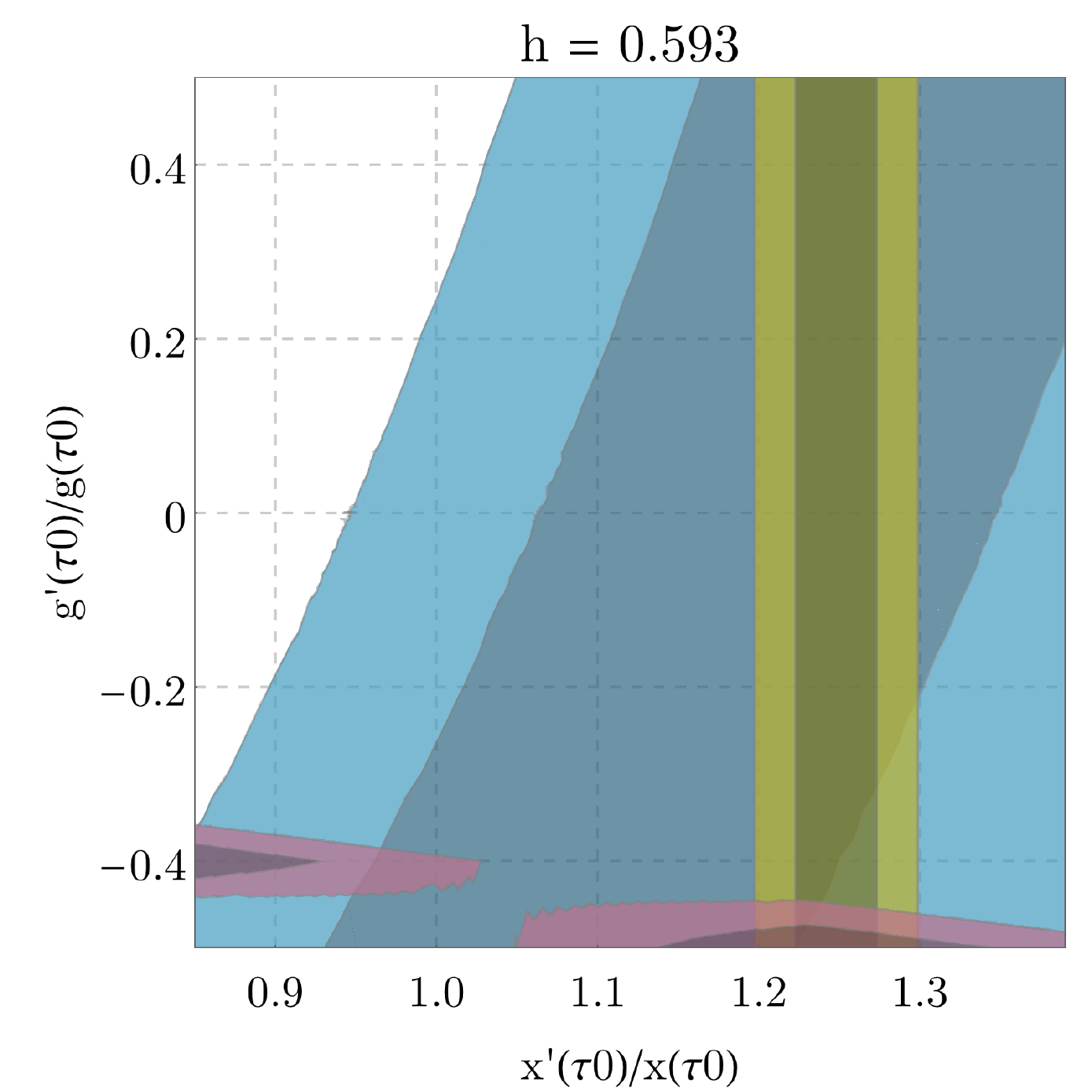}
  \includegraphics[trim={10 1 10 6},clip,width=.32\linewidth]{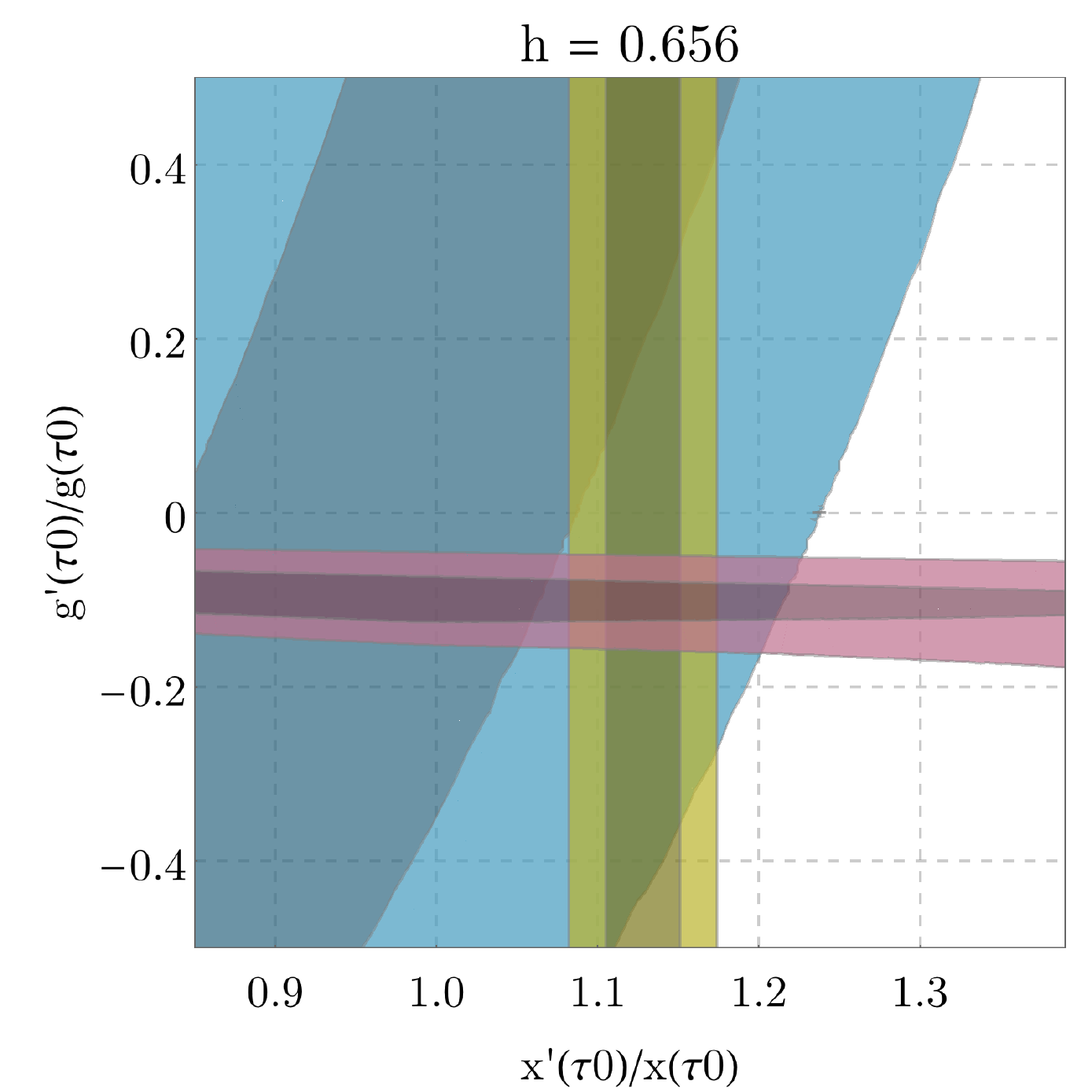}
  \includegraphics[trim={10 1 10 6},clip,width=.32\linewidth]{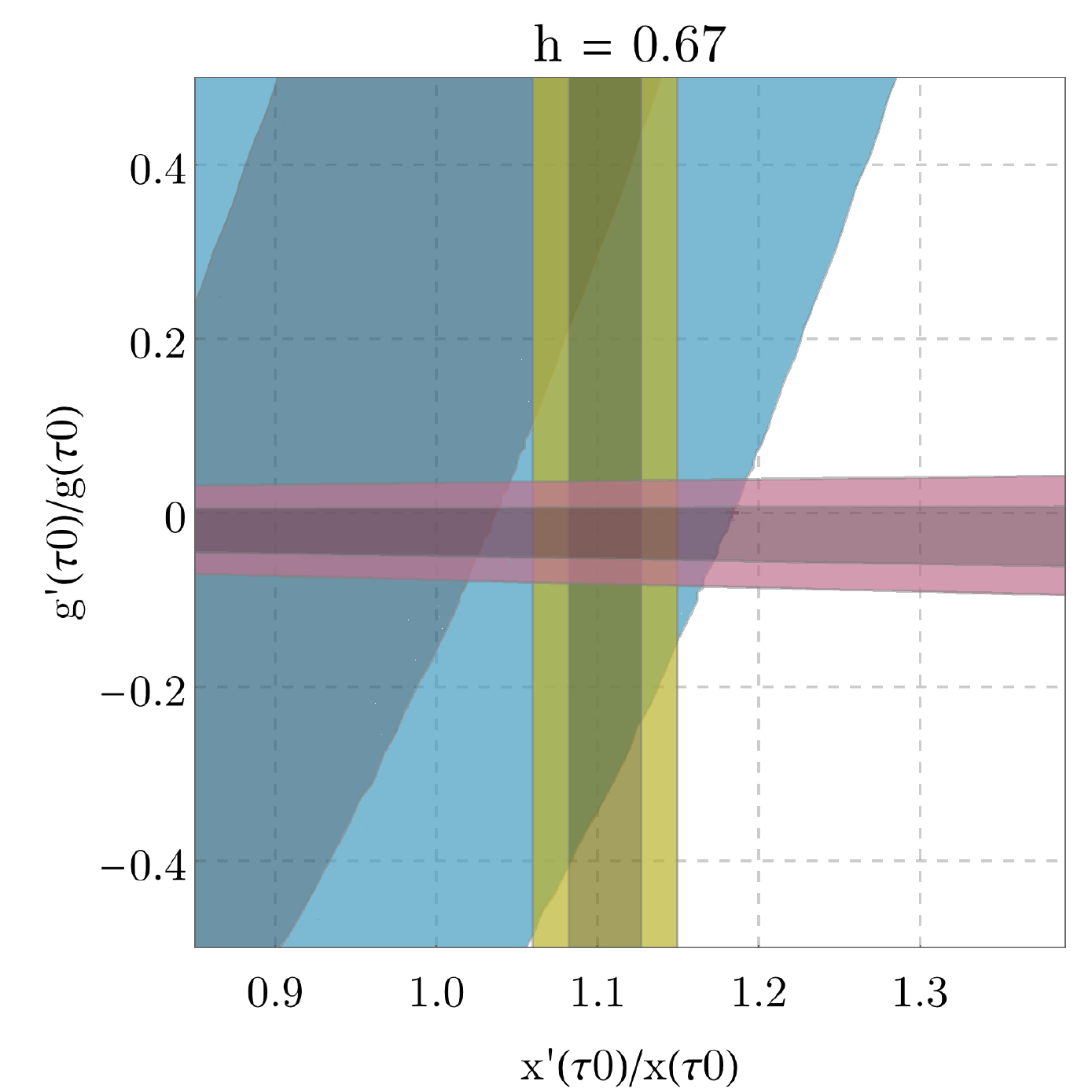}\\
  \includegraphics[trim={10 1 10 6},clip,width=.32\linewidth]{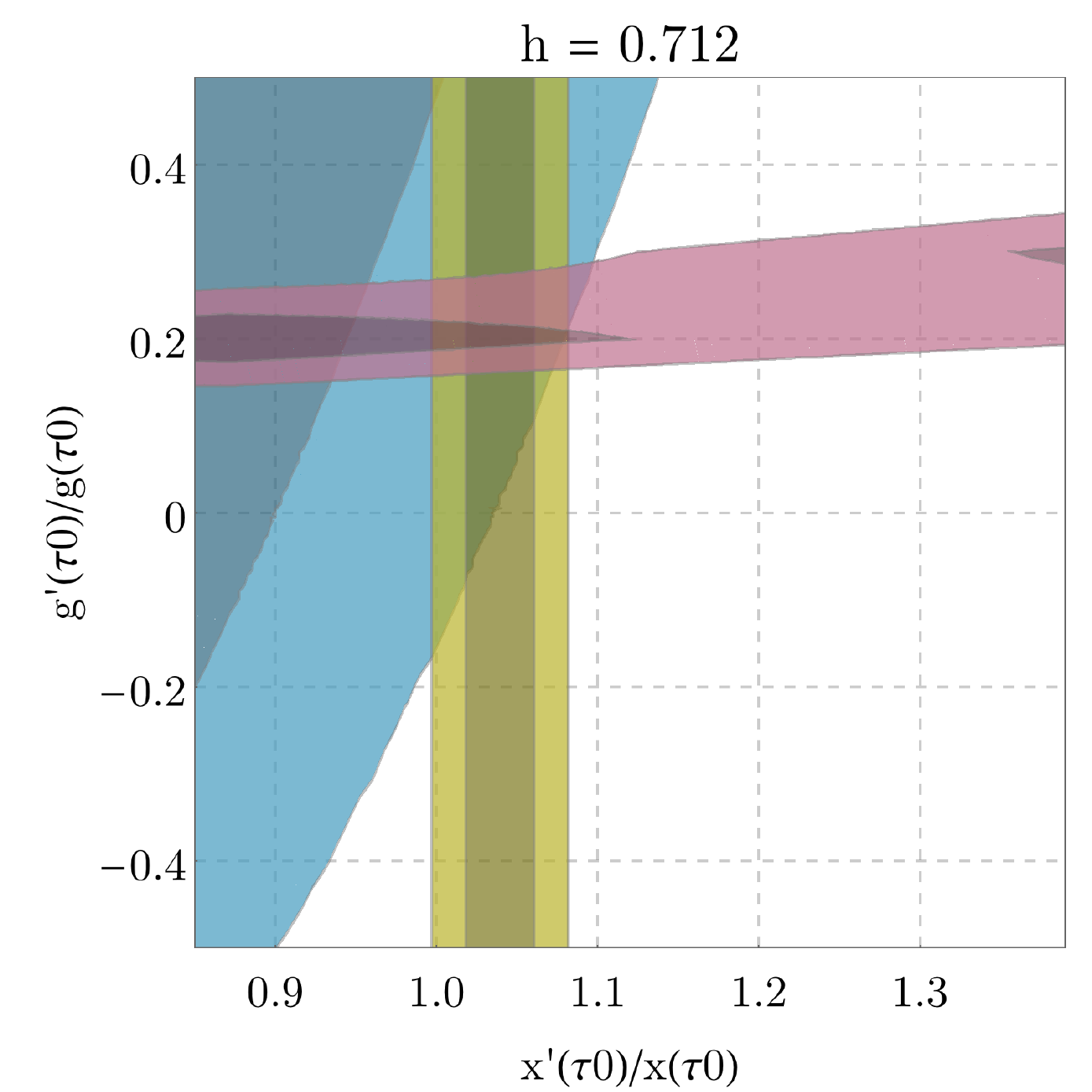}
  \includegraphics[trim={10 1 10 6},clip,width=.32\linewidth]{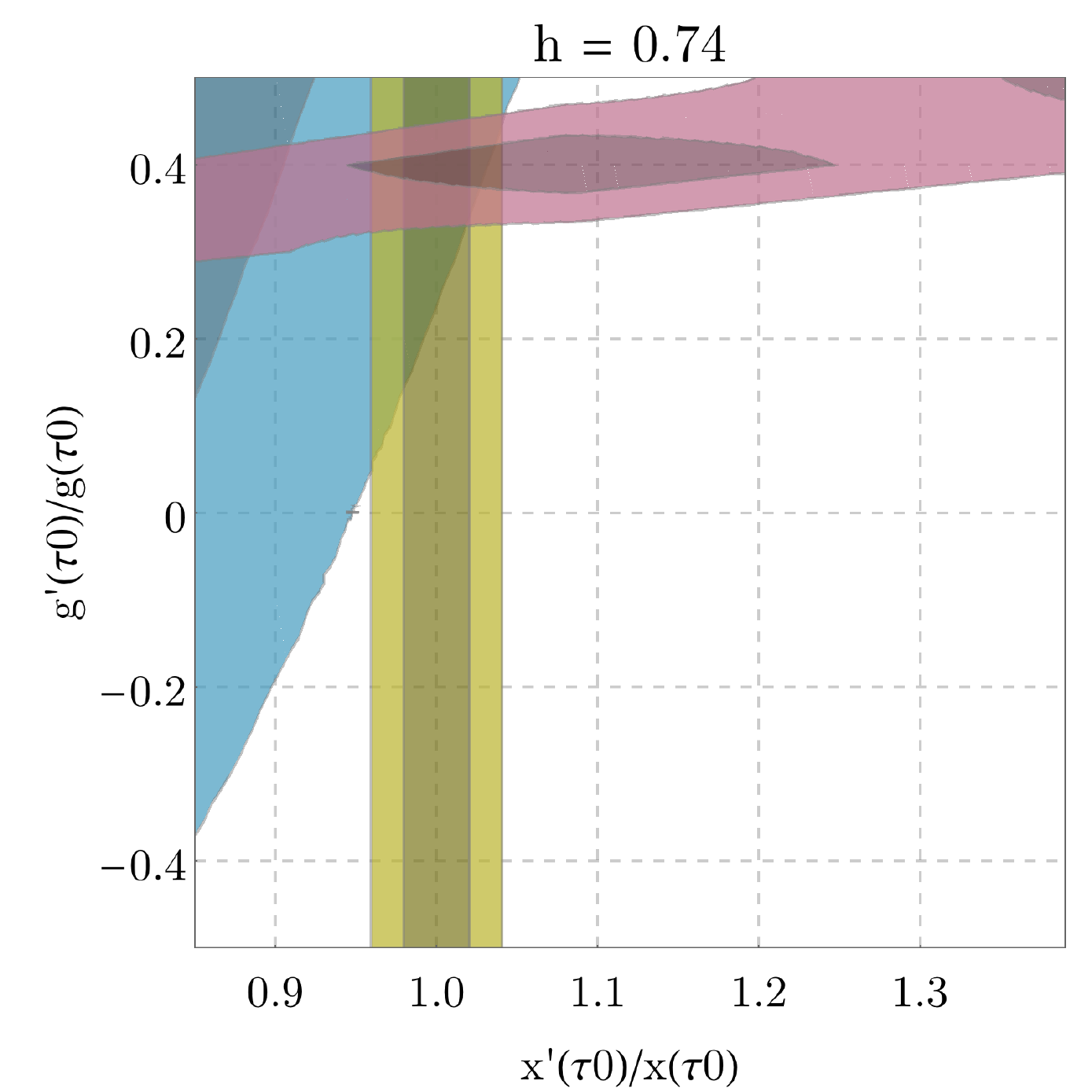}
\caption{Contour plots of $\Sigma < 1$ (darker color) and $\Sigma < 2$ (lighter color) for fixed values of $h$. Further comments in the text.}
\label{regions2}
\end{figure}

In figure \ref{regions2} we show contour plots of $\Sigma < 1,2$ regions in the $(\alpha,\gamma)$ plane for fixed values of $h$. 
The plot for $h=0.67$ shows cases of the scale-dependent cosmology that could contain the $\LCDM$ model in the sense that the $|\gamma| \rightarrow 0$ region is within the $\Sigma_{(ii)}<1$ region. This means that for sufficiently small values of $\gamma$ both models will, within the given accuracy, predict the same values for the parameters. 
The point $\gamma=0 $ is special because the scale-dependent model degenerates to the $\LCDM$ model in the sense that $\gamma$, $\gamma'$ and all the higher-order derivatives of $\gamma$ become exactly zero. This mathematical fact can be seen by taking derivatives of $\gamma$ and using the NEC recursively in order to express terms containing $g''/g$ as a product of $\gamma$ and a complicated polynomial of $\alpha$, $\gamma$, $\alpha'$, $\gamma'$, ...,$\alpha^{(n-1)}$, $\gamma^{(n-1)}$,
\begin{align}
    \gamma'=&\gamma (\alpha+\gamma)\,,\label{recur}\\
    \gamma''=&\gamma ((\alpha+\gamma)^2+\alpha'+\gamma')\,, \\
    \vdots \nonumber\\
    \gamma^{(n)}=& \gamma \times \text{Pol}(\alpha, \gamma, \alpha', \gamma', \cdots,\alpha^{(n-1)}, \gamma^{(n-1)})\,.
\end{align}
This mathematical property of the SD model presented here can be used to justify the NEC (eq. \ref{NECCosm}) as a meaningful condition that allow us to construct a generalization of $\LCDM$ such that the general model still contains $\LCDM$ in a region of the phase space of the model. The argument can therefore be reverted to use the smallness of $\gamma$ (at present) to understand the great success of $\LCDM$.

In figure \ref{regions2} we have also two plots for smaller values $h=0.593,0.656$ and two plots for bigger values $h=0.712,0.74$. All plots of figure \ref{regions2} show intersections of regions of $\Sigma_{(i)}<1$, $\Sigma_{(ii)}<1$ and $\Sigma_{(iii)}<2$. A triple intersection $\Sigma_{(i)},\Sigma_{(ii)},\Sigma_{(iii)}<1$ can be found in the $h=0.593$ plot for values of $(\alpha,\gamma)$ that are consistent with the plot for $\gamma = -0.6$ of figure \ref{regions1}.

A complete analysis should infer the preferred values for the density parameters within the context of the scale-dependent cosmology, however, this is beyond the scope of the present paper.

%
\section{Conclusions and final remarks}\label{conlusion}

We have discussed an extension of the $\LCDM$ model that is based on a scale--dependent generalization of gravity.
Such a modification is well motivated 
by any effective quantum gravity with 
possible infrared instabilities, as for example
asymptotically safe quantum gravity.
In contrast to previous analytical approaches, where we could control the scale--dependent effect employing the scale--dependent parameter $\epsilon$, we now control the same effect via the initial conditions on the 
dynamical system (\ref{eq1}-\ref{eq3}).

We showed that the resulting, 
very restricted, form of scale-dependence of Newton's and the cosmological coupling
gives promising results,
alleviating 
the well-known discrepancy between early and
late time measurements of the Hubble parameter $H_0$.
We performed a numerical scan over the phase space of the model. In order to pin down initial conditions, we demanded compatibility with the CMB-inferred parameters, late time measurements of $H_0$, and the age of the
Universe. We collected these results as contour plots, where the above-mentioned criteria are contrasted. We also commented on the possible degeneracy with the $\LCDM$ model when $\dot{G}/G$ is small enough, see plot for $h=0.67$ in figure \ref{regions2}.

Neglecting higher-order corrections to the CMB-inferred parameters, we found plenty of regions where tension on early and late time measurements of $H_0$ can be released. 
This works even for values of $\dot{G}/G \sim 10^ {-14}$ which is in compliance with the most strict constraints on $\dot{G}(t)$, see plots of second line in figure \ref{regions2}. Values of $\dot{G}/G \sim 10^{-12}$, that are in the upper region of existing cosmological constraints on $\dot{G}(t)$ can remove the tension with special ease, see plots of the first line in figure \ref{regions2}.

As next step, it is planned to study perturbations to find the best-fit parameters to generate the CMB structure based on the scale--dependent cosmological model presented in this paper.
This would be particularly interesting
for scenarios which go beyond the weak scale--dependence hypothesis discussed in
subsection \ref{perturbations}.
Based on our findings, we expect small corrections to the inferred values from $\Lambda$CDM and compatibility with the late cosmological time measurements of $H_0$.
Further future developments will be the computation of cosmological constraints on $\dot{G}/G$ based on the scale--dependent cosmological model and the comparison to other cosmological models beyond $\Lambda$CDM by using the statefinder parameters \cite{Panotopoulos:2018sso,Panotopoulos:2019xbw,Panotopoulos:2017mte,Zimdahl:2003wg,Sahni:2002fz,Panotopoulos:2007zn}. 
We also plan to explore how the results change when the system (\ref{SD1}-\ref{SD2}) is completed by other relations, alternative
to~(\ref{nec}).

\section*{Acknowlegements}

We are grateful to the anonymous reviewer for a careful
reading of the manuscript as well as for numerous
useful comments and suggestions, which helped us significantly improve the quality of our work.
The author A. R. acknowledges DI-VRIEA for financial support through Proyecto Postdoctorado 2019 VRIEA-PUCV. P. A. acknowledge MINEDUC-UA project ANT 1755 and Semillero de Investigación project SEM18-02 from Universidad de Antofagasta, Chile.


\begin{thebibliography}{99}

\bibitem{Aghanim:2018eyx}
N.~Aghanim \textit{et al.} [Planck],
Astron. Astrophys. \textbf{641} (2020), A6
doi:10.1051/0004-6361/201833910
[arXiv:1807.06209 [astro-ph.CO]].

\bibitem{Ade:2015rim}
P.~A.~R.~Ade \textit{et al.} [Planck],
Astron. Astrophys. \textbf{594} (2016), A14
doi:10.1051/0004-6361/201525814
[arXiv:1502.01590 [astro-ph.CO]].

\bibitem{DiValentino:2019dzu}
E.~Di Valentino, A.~Melchiorri and J.~Silk,
JCAP \textbf{01} (2020), 013
doi:10.1088/1475-7516/2020/01/013
[arXiv:1908.01391 [astro-ph.CO]].

\bibitem{Akrami:2018odb}
Y.~Akrami \textit{et al.} [Planck],
Astron. Astrophys. \textbf{641} (2020), A10
doi:10.1051/0004-6361/201833887
[arXiv:1807.06211 [astro-ph.CO]].

\bibitem{Riess:2019cxk}
A.~G.~Riess, S.~Casertano, W.~Yuan, L.~M.~Macri and D.~Scolnic,
Astrophys. J. \textbf{876} (2019) no.1, 85
doi:10.3847/1538-4357/ab1422
[arXiv:1903.07603 [astro-ph.CO]].

\bibitem{Verde:2019ivm}
L.~Verde, T.~Treu and A.~G.~Riess,
doi:10.1038/s41550-019-0902-0
[arXiv:1907.10625 [astro-ph.CO]].

\bibitem{Aylor:2018drw}
K.~Aylor, M.~Joy, L.~Knox, M.~Millea, S.~Raghunathan and W.~L.~K.~Wu,
Astrophys. J. \textbf{874} (2019) no.1, 4
doi:10.3847/1538-4357/ab0898
[arXiv:1811.00537 [astro-ph.CO]].

\bibitem{Knox:2019rjx}
L.~Knox and M.~Millea,
Phys. Rev. D \textbf{101} (2020) no.4, 043533
doi:10.1103/PhysRevD.101.043533
[arXiv:1908.03663 [astro-ph.CO]].

\bibitem{Poulin:2018cxd}
V.~Poulin, T.~L.~Smith, T.~Karwal and M.~Kamionkowski,
Phys. Rev. Lett. \textbf{122} (2019) no.22, 221301
doi:10.1103/PhysRevLett.122.221301
[arXiv:1811.04083 [astro-ph.CO]].

\bibitem{Agrawal:2019lmo}
P.~Agrawal, F.~Y.~Cyr-Racine, D.~Pinner and L.~Randall,
[arXiv:1904.01016 [astro-ph.CO]].

\bibitem{Kreisch:2019yzn}
C.~D.~Kreisch, F.~Y.~Cyr-Racine and O.~Dor\'e,
Phys. Rev. D \textbf{101} (2020) no.12, 123505
doi:10.1103/PhysRevD.101.123505
[arXiv:1902.00534 [astro-ph.CO]].

\bibitem{Alestas:2020mvb}
G.~Alestas, L.~Kazantzidis and L.~Perivolaropoulos,
Phys. Rev. D \textbf{101} (2020) no.12, 123516
doi:10.1103/PhysRevD.101.123516
[arXiv:2004.08363 [astro-ph.CO]].

\bibitem{Houthoff:2017oam}
W.~B.~Houthoff, A.~Kurov and F.~Saueressig,
Eur. Phys. J. C \textbf{77} (2017), 491
doi:10.1140/epjc/s10052-017-5046-8
[arXiv:1705.01848 [hep-th]].

\bibitem{Wetterich:2018qsl}
C.~Wetterich,
Phys. Rev. D \textbf{98} (2018) no.2, 026028
doi:10.1103/PhysRevD.98.026028
[arXiv:1802.05947 [gr-qc]].

\bibitem{Bosma:2019aiu}
L.~Bosma, B.~Knorr and F.~Saueressig,
Phys. Rev. Lett. \textbf{123} (2019) no.10, 101301
doi:10.1103/PhysRevLett.123.101301
[arXiv:1904.04845 [hep-th]].

\bibitem{Bonanno:2017pkg}
A.~Bonanno and F.~Saueressig,
Comptes Rendus Physique \textbf{18} (2017), 254-264
doi:10.1016/j.crhy.2017.02.002
[arXiv:1702.04137 [hep-th]].

\bibitem{Reuter:2012xf}
M.~Reuter and F.~Saueressig,
Lect. Notes Phys. \textbf{863} (2013), 185-223
doi:10.1007/978-3-642-33036-0\_8
[arXiv:1205.5431 [hep-th]].

\bibitem{Bonanno:2001xi}
A.~Bonanno and M.~Reuter,
Phys. Rev. D \textbf{65} (2002), 043508
doi:10.1103/PhysRevD.65.043508
[arXiv:hep-th/0106133 [hep-th]].

\bibitem{Lima:1994gi}
J.~A.~S.~Lima and J.~M.~F.~Maia,
Phys. Rev. D \textbf{49} (1994), 5597-5600
doi:10.1103/PhysRevD.49.5597

\bibitem{Lima:1995ea}
J.~A.~S.~Lima and M.~Trodden,
Phys. Rev. D \textbf{53} (1996), 4280-4286
doi:10.1103/PhysRevD.53.4280
[arXiv:astro-ph/9508049 [astro-ph]].

\bibitem{Weinberg:1988cp}
S.~Weinberg,
Rev. Mod. Phys. \textbf{61} (1989), 1-23
doi:10.1103/RevModPhys.61.1

\bibitem{Sahni:1999gb}
V.~Sahni and A.~A.~Starobinsky,
Int. J. Mod. Phys. D \textbf{9} (2000), 373-444
doi:10.1142/S0218271800000542
[arXiv:astro-ph/9904398 [astro-ph]].

\bibitem{Padmanabhan:2002ji}
T.~Padmanabhan,
Phys. Rept. \textbf{380} (2003), 235-320
doi:10.1016/S0370-1573(03)00120-0
[arXiv:hep-th/0212290 [hep-th]].

\bibitem{Peebles:2002gy}
P.~J.~E.~Peebles and B.~Ratra,
Rev. Mod. Phys. \textbf{75} (2003), 559-606
doi:10.1103/RevModPhys.75.559
[arXiv:astro-ph/0207347 [astro-ph]].

\bibitem{Sola:2013gha}
J.~Sola,
J. Phys. Conf. Ser. \textbf{453} (2013), 012015
doi:10.1088/1742-6596/453/1/012015
[arXiv:1306.1527 [gr-qc]].

\bibitem{Koch:2010nn}
B.~Koch and I.~Ramirez,
Class. Quant. Grav. \textbf{28} (2011), 055008
doi:10.1088/0264-9381/28/5/055008
[arXiv:1010.2799 [gr-qc]].

\bibitem{Contreras:2013hua}
C.~Contreras, B.~Koch and P.~Rioseco,
Class. Quant. Grav. \textbf{30} (2013), 175009
doi:10.1088/0264-9381/30/17/175009
[arXiv:1303.3892 [astro-ph.CO]].

\bibitem{Koch:2013rwa}
B.~Koch, C.~Contreras, P.~Rioseco and F.~Saueressig,
Springer Proc. Phys. \textbf{170} (2016), 263-269
doi:10.1007/978-3-319-20046-0\_31
[arXiv:1311.1121 [hep-th]].

\bibitem{Koch:2014joa}
B.~Koch, P.~Rioseco and C.~Contreras,
Phys. Rev. D \textbf{91} (2015) no.2, 025009
doi:10.1103/PhysRevD.91.025009
[arXiv:1409.4443 [hep-th]].

\bibitem{Koch:2015nva}
B.~Koch and P.~Rioseco,
Class. Quant. Grav. \textbf{33} (2016), 035002
doi:10.1088/0264-9381/33/3/035002
[arXiv:1501.00904 [gr-qc]].

\bibitem{Koch:2016uso}
B.~Koch, I.~A.~Reyes and \'A.~Rinc\'on,
Class. Quant. Grav. \textbf{33} (2016) no.22, 225010
doi:10.1088/0264-9381/33/22/225010
[arXiv:1606.04123 [hep-th]].

\bibitem{Rincon:2017ypd}
\'A.~Rinc\'on, B.~Koch and I.~Reyes,
J. Phys. Conf. Ser. \textbf{831} (2017) no.1, 012007
doi:10.1088/1742-6596/831/1/012007
[arXiv:1701.04531 [hep-th]].

\bibitem{Rincon:2017goj}
\'A.~Rinc\'on, E.~Contreras, P.~Bargue\~no, B.~Koch, G.~Panotopoulos and A.~Hern\'andez-Arboleda,
Eur. Phys. J. C \textbf{77} (2017) no.7, 494
doi:10.1140/epjc/s10052-017-5045-9
[arXiv:1704.04845 [hep-th]].

\bibitem{Rincon:2017ayr}
\'A.~Rinc\'on and B.~Koch,
J. Phys. Conf. Ser. \textbf{1043} (2018) no.1, 012015
doi:10.1088/1742-6596/1043/1/012015
[arXiv:1705.02729 [hep-th]].

\bibitem{Contreras:2017eza}
E.~Contreras, \'A.~Rinc\'on, B.~Koch and P.~Bargue\~no,
Int. J. Mod. Phys. D \textbf{27} (2017) no.03, 1850032
doi:10.1142/S0218271818500323
[arXiv:1711.08400 [gr-qc]].

\bibitem{Rincon:2018sgd}
\'A.~Rinc\'on and G.~Panotopoulos,
Phys. Rev. D \textbf{97} (2018) no.2, 024027
doi:10.1103/PhysRevD.97.024027
[arXiv:1801.03248 [hep-th]].

\bibitem{Hernandez-Arboleda:2018qdo}
A.~Hern\'andez-Arboleda, \'A.~Rinc\'on, B.~Koch, E.~Contreras and P.~Bargue\~no,
[arXiv:1802.05288 [gr-qc]].

\bibitem{Contreras:2018dhs}
E.~Contreras, \'A.~Rinc\'on, B.~Koch and P.~Bargue\~no,
Eur. Phys. J. C \textbf{78} (2018) no.3, 246
doi:10.1140/epjc/s10052-018-5709-0
[arXiv:1803.03255 [gr-qc]].

\bibitem{Rincon:2018lyd}
\'A.~Rinc\'on and B.~Koch,
Eur. Phys. J. C \textbf{78} (2018) no.12, 1022
doi:10.1140/epjc/s10052-018-6488-3
[arXiv:1806.03024 [hep-th]].

\bibitem{Rincon:2018dsq}
\'A.~Rinc\'on, E.~Contreras, P.~Bargue\~no, B.~Koch and G.~Panotopoulos,
Eur. Phys. J. C \textbf{78} (2018) no.8, 641
doi:10.1140/epjc/s10052-018-6106-4
[arXiv:1807.08047 [hep-th]].

\bibitem{Contreras:2018gct}
E.~Contreras, \'A.~Rinc\'on and J.~M.~Ram\'\i{}rez-Velasquez,
Eur. Phys. J. C \textbf{79} (2019) no.1, 53
doi:10.1140/epjc/s10052-019-6601-2
[arXiv:1810.07356 [gr-qc]].

\bibitem{Canales:2018tbn}
F.~Canales, B.~Koch, C.~Laporte and \'A.~Rinc\'on,
JCAP \textbf{01} (2020), 021
doi:10.1088/1475-7516/2020/01/021
[arXiv:1812.10526 [gr-qc]].

\bibitem{Rincon:2019cix}
\'A.~Rinc\'on, E.~Contreras, P.~Bargue\~no and B.~Koch,
Eur. Phys. J. Plus \textbf{134} (2019) no.11, 557
doi:10.1140/epjp/i2019-13081-5
[arXiv:1901.03650 [gr-qc]].

\bibitem{Rincon:2019zxk}
\'A.~Rinc\'on and J.~R.~Villanueva,
Class. Quant. Grav. \textbf{37} (2020) no.17, 175003
doi:10.1088/1361-6382/aba17f
[arXiv:1902.03704 [gr-qc]].

\bibitem{Contreras:2019fwu}
E.~Contreras, \'A.~Rinc\'on and P.~Bargue\~no,
Eur. Phys. J. C \textbf{80} (2020) no.5, 367
doi:10.1140/epjc/s10052-020-7936-4
[arXiv:1902.05941 [gr-qc]].

\bibitem{Fathi:2019jid}
M.~Fathi, \'A.~Rinc\'on and J.~R.~Villanueva,
Class. Quant. Grav. \textbf{37} (2020) no.7, 075004
doi:10.1088/1361-6382/ab6f7c
[arXiv:1903.09037 [gr-qc]].

\bibitem{Rincon:2019ptp}
\'A.~Rinc\'on,
''Black holes in scale-dependent frameworks,''

\bibitem{Contreras:2019cmf}
E.~Contreras, \'A.~Rinc\'on, G.~Panotopoulos, P.~Bargue\~no and B.~Koch,
Phys. Rev. D \textbf{101} (2020) no.6, 064053
doi:10.1103/PhysRevD.101.064053
[arXiv:1906.06990 [gr-qc]].

\bibitem{Panotopoulos:2020zqa}
G.~Panotopoulos, \'A.~Rinc\'on and I.~Lopes,
Eur. Phys. J. C \textbf{80} (2020) no.4, 318
doi:10.1140/epjc/s10052-020-7900-3
[arXiv:2004.02627 [gr-qc]].

\bibitem{Panotopoulos:2021tkk} 
  G.~Panotopoulos, A.~Rincón and I.~Lopes,
  Phys.\ Rev.\ D {\bf 103}, 104040 (2021)
  [arXiv:2104.13611 [gr-qc]].

\bibitem{Panotopoulos:2021obe} 
  G.~Panotopoulos, Á.~Rincón and I.~Lopes,
  Eur.\ Phys.\ J.\ C {\bf 81}, no. 1, 63 (2021)
  [arXiv:2101.06649 [gr-qc]].

\bibitem{Rincon:2021hjj} 
  Á.~Rincón, E.~Contreras, P.~Bargueño, B.~Koch and G.~Panotopoulos,
  Phys.\ Dark Univ.\  {\bf 31}, 100783 (2021)
  [arXiv:2102.02426 [gr-qc]].


\bibitem{Kontou:2020bta}
E.~A.~Kontou and K.~Sanders,
Class. Quant. Grav. \textbf{37} (2020) no.19, 193001
doi:10.1088/1361-6382/ab8fcf
[arXiv:2003.01815 [gr-qc]].

\bibitem{Penrose:1964wq}
R.~Penrose,
Phys. Rev. Lett. \textbf{14} (1965), 57-59
doi:10.1103/PhysRevLett.14.57

\bibitem{Hawking:1969sw}
S.~W.~Hawking and R.~Penrose,
Proc. Roy. Soc. Lond. A \textbf{314} (1970), 529-548
doi:10.1098/rspa.1970.0021

\bibitem{Wall:2009wi}
A.~C.~Wall,
Phys. Rev. D \textbf{81} (2010), 024038
doi:10.1103/PhysRevD.81.024038
[arXiv:0910.5751 [gr-qc]].

\bibitem{Parikh:2014mja}
M.~Parikh and J.~P.~van der Schaar,
Phys. Rev. D \textbf{91} (2015) no.8, 084002
doi:10.1103/PhysRevD.91.084002
[arXiv:1406.5163 [hep-th]].

\bibitem{Epstein:1965zza}
H.~Epstein, V.~Glaser and A.~Jaffe,
Nuovo Cim. \textbf{36} (1965), 1016
doi:10.1007/BF02749799

\bibitem{Visser:1999de}
M.~Visser and C.~Barcelo,
doi:10.1142/9789812792129\_0014
[arXiv:gr-qc/0001099 [gr-qc]].

\bibitem{Barcelo:2002bv}
C.~Barcelo and M.~Visser,
Int. J. Mod. Phys. D \textbf{11} (2002), 1553-1560
doi:10.1142/S0218271802002888
[arXiv:gr-qc/0205066 [gr-qc]].

\bibitem{Arefeva:2006ido}
I.~Y.~Aref'eva and I.~V.~Volovich,
Theor. Math. Phys. \textbf{155} (2008), 503-511
doi:10.1007/s11232-008-0041-8
[arXiv:hep-th/0612098 [hep-th]].

\bibitem{Curiel:2014zba}
E.~Curiel,
Einstein Stud. \textbf{13} (2017), 43-104
doi:10.1007/978-1-4939-3210-8\_3
[arXiv:1405.0403 [physics.hist-ph]].

\bibitem{Ford:1978qya}
L.~H.~Ford,
Proc. Roy. Soc. Lond. A \textbf{364} (1978), 227-236
doi:10.1098/rspa.1978.0197

\bibitem{Fewster:2010gm}
C.~J.~Fewster and G.~J.~Galloway,
Class. Quant. Grav. \textbf{28} (2011), 125009
doi:10.1088/0264-9381/28/12/125009
[arXiv:1012.6038 [gr-qc]].

\bibitem{Ecker:2017jdw}
C.~Ecker, D.~Grumiller, W.~van der Schee and P.~Stanzer,
Phys. Rev. D \textbf{97} (2018) no.12, 126016
doi:10.1103/PhysRevD.97.126016
[arXiv:1710.09837 [hep-th]].

\bibitem{Grumiller:2019xna}
D.~Grumiller, P.~Parekh and M.~Riegler,
Phys. Rev. Lett. \textbf{123} (2019) no.12, 121602
doi:10.1103/PhysRevLett.123.121602
[arXiv:1907.06650 [hep-th]].

\bibitem{Ecker:2019ocp}
C.~Ecker, D.~Grumiller, W.~van der Schee, M.~M.~Sheikh-Jabbari and P.~Stanzer,
SciPost Phys. \textbf{6} (2019) no.3, 036
doi:10.21468/SciPostPhys.6.3.036
[arXiv:1901.04499 [hep-th]].

\bibitem{Bonanos}
S. Bonanos,
http://www-old.inp.demokritos.gr/~sbonano//RGTC/



\bibitem{Fixsen:2009ug}
D.~J.~Fixsen,
Astrophys. J. \textbf{707} (2009), 916-920
doi:10.1088/0004-637X/707/2/916
[arXiv:0911.1955 [astro-ph.CO]].

\bibitem{Reuter:2003ca}
M.~Reuter and H.~Weyer,
Phys. Rev. D \textbf{69} (2004), 104022
doi:10.1103/PhysRevD.69.104022
[arXiv:hep-th/0311196 [hep-th]].

\bibitem{Domazet:2012tw}
S.~Domazet and H.~Stefancic,
Class. Quant. Grav. \textbf{29} (2012), 235005
doi:10.1088/0264-9381/29/23/235005
[arXiv:1204.1483 [gr-qc]].

\bibitem{Contreras:2016mdt}
C.~Contreras, B.~Koch and P.~Rioseco,
J. Phys. Conf. Ser. \textbf{720} (2016) no.1, 012020
doi:10.1088/1742-6596/720/1/012020

\bibitem{Chauvineau:2015cha}
B.~Chauvineau, D.~C.~Rodrigues and J.~C.~Fabris,
Gen. Rel. Grav. \textbf{48} (2016) no.6, 80
doi:10.1007/s10714-016-2075-9
[arXiv:1503.07581 [gr-qc]].

\bibitem{Carroll:1997ar}
S.~M.~Carroll,
[arXiv:gr-qc/9712019 [gr-qc]].

\bibitem{Giani:2019vjf}
L.~Giani, T.~Miranda and O.~F.~Piattella,
Phys. Dark Univ. \textbf{26} (2019), 100357
[arXiv:1905.02720 [gr-qc]].

\bibitem{Sola:2019jek}
J.~Sol\`a Peracaula, A.~Gomez-Valent, J.~de Cruz P\'erez and C.~Moreno-Pulido,
Astrophys. J. Lett. \textbf{886} (2019) no.1, L6
doi:10.3847/2041-8213/ab53e9
[arXiv:1909.02554 [astro-ph.CO]].

\bibitem{Sola:2020lba}
J.~Sola, A.~Gomez-Valent, J.~d.~Perez and C.~Moreno-Pulido,
doi:10.1088/1361-6382/abbc43
[arXiv:2006.04273 [astro-ph.CO]].

\bibitem{Joudaki:2020shz}
S.~Joudaki, P.~G.~Ferreira, N.~A.~Lima and H.~A.~Winther,
[arXiv:2010.15278 [astro-ph.CO]].

\bibitem{Will:2014kxa}
C.~M.~Will,
Living Rev. Rel. \textbf{17} (2014), 4
doi:10.12942/lrr-2014-4
[arXiv:1403.7377 [gr-qc]].

\bibitem{Bardeen:1980kt}
J.~M.~Bardeen,
Phys. Rev. D \textbf{22} (1980), 1882-1905
doi:10.1103/PhysRevD.22.1882

\bibitem{Fry:1983cj}
J.~N.~Fry,
Astrophys. J. \textbf{279} (1984), 499-510
doi:10.1086/161913

\bibitem{Kodama:1985bj}
H.~Kodama and M.~Sasaki,
Prog. Theor. Phys. Suppl. \textbf{78} (1984), 1-166
doi:10.1143/PTPS.78.1

\bibitem{Toniato:2017wmk}
J.~D.~Toniato, D.~C.~Rodrigues, \'A.~O.~F.~de Almeida and N.~Bertini,
Phys. Rev. D \textbf{96} (2017) no.6, 064034
doi:10.1103/PhysRevD.96.064034
[arXiv:1706.09032 [gr-qc]].

\bibitem{Uzan:2002vq}
J.~P.~Uzan,
Rev. Mod. Phys. \textbf{75} (2003), 403
doi:10.1103/RevModPhys.75.403
[arXiv:hep-ph/0205340 [hep-ph]].

\bibitem{Uzan:2010pm}
J.~P.~Uzan,
Living Rev. Rel. \textbf{14} (2011), 2
doi:10.12942/lrr-2011-2
[arXiv:1009.5514 [astro-ph.CO]].

\bibitem{Williams:2012nc}
J.~G.~Williams, S.~G.~Turyshev and D.~Boggs,
Class. Quant. Grav. \textbf{29} (2012), 184004
doi:10.1088/0264-9381/29/18/184004
[arXiv:1203.2150 [gr-qc]].

\bibitem{Hofmann:2018myc}
F.~Hofmann and J.~M\"uller,
Class. Quant. Grav. \textbf{35} (2018) no.3, 035015
doi:10.1088/1361-6382/aa8f7a

\bibitem{Fienga:2014bvy}
A.~Fienga, J.~Laskar, P.~Exertier, H.~Manche and M.~Gastineau,
[arXiv:1409.4932 [astro-ph.EP]].

\bibitem{Pitjeva:2013xxa}
E.~V.~Pitjeva and N.~P.~Pitjev,
Mon. Not. Roy. Astron. Soc. \textbf{432} (2013), 3431
doi:10.1093/mnras/stt695
[arXiv:1306.3043 [astro-ph.EP]].

\bibitem{Deller:2008jx}
A.~T.~Deller, J.~P.~W.~Verbiest, S.~J.~Tingay and M.~Bailes,
Astrophys. J. Lett. \textbf{685} (2008), L67
doi:10.1086/592401
[arXiv:0808.1594 [astro-ph]].

\bibitem{Zhu:2018etc}
W.~W.~Zhu, G.~Desvignes, N.~Wex, R.~N.~Caballero, D.~J.~Champion, P.~B.~Demorest, J.~A.~Ellis, G.~H.~Janssen, M.~Kramer and A.~Krieger, \textit{et al.}
Mon. Not. Roy. Astron. Soc. \textbf{482} (2019) no.3, 3249-3260
doi:10.1093/mnras/sty2905
[arXiv:1802.09206 [astro-ph.HE]].


\bibitem{Genova:2018gti}
A.~Genova, E.~Mazarico, S.~Goossens, F.~G.~Lemoine, G.~A.~Neumann, D.~E.~Smith and M.~T.~Zuber,
Nature Communications. \textbf{289} (2018) no.9
doi:10.1038/s41467-017-02558-1


\bibitem{Masuda:2016ggi}
K.~Masuda and Y.~Suto,
Publ. Astron. Soc. Jap. \textbf{68} (2016) no.3, L5
doi:10.1093/pasj/psw017
[arXiv:1602.02513 [gr-qc]].

\bibitem{Jofre:2006ug}
P.~Jofre, A.~Reisenegger and R.~Fernandez,
Phys. Rev. Lett. \textbf{97} (2006), 131102
doi:10.1103/PhysRevLett.97.131102
[arXiv:astro-ph/0606708 [astro-ph]].

\bibitem{Corsico:2013ida}
A.~H.~C\'orsico, L.~G.~Althaus, E.~Garc\'\i{}a-Berro and A.~D.~Romero,
JCAP \textbf{06} (2013), 032
doi:10.1088/1475-7516/2013/06/032
[arXiv:1306.1864 [astro-ph.SR]].

\bibitem{Copi:2003xd}
C.~J.~Copi, A.~N.~Davis and L.~M.~Krauss,
Phys. Rev. Lett. \textbf{92} (2004), 171301
doi:10.1103/PhysRevLett.92.171301
[arXiv:astro-ph/0311334 [astro-ph]].

\bibitem{Guenther:2009xp}
D.~B.~Guenther, L.~M.~Krauss and P.~Demarque,
The Astrophysical Journal. \textbf{92} (1998)
doi:10.1086/305567

\bibitem{Wu:2009zb}
F.~Wu and X.~Chen,
Phys. Rev. D \textbf{82} (2010), 083003
doi:10.1103/PhysRevD.82.083003
[arXiv:0903.0385 [astro-ph.CO]].

\bibitem{Yunes:2009bv}
N.~Yunes, F.~Pretorius and D.~Spergel,
Phys. Rev. D \textbf{81} (2010), 064018
doi:10.1103/PhysRevD.81.064018
[arXiv:0912.2724 [gr-qc]].

\bibitem{Zhao:2018gwk}
W.~Zhao, B.~S.~Wright and B.~Li,
JCAP \textbf{10} (2018), 052
doi:10.1088/1475-7516/2018/10/052
[arXiv:1804.03066 [astro-ph.CO]].

\bibitem{Freedman:2000cf}
W.~L.~Freedman \textit{et al.} [HST],
Astrophys. J. \textbf{553} (2001), 47-72
doi:10.1086/320638
[arXiv:astro-ph/0012376 [astro-ph]].

\bibitem{Bond:2013jka}
H.~E.~Bond, E.~P.~Nelan, D.~A.~VandenBerg, G.~H.~Schaefer and D.~Harmer,
Astrophys. J. Lett. \textbf{765} (2013), L12
doi:10.1088/2041-8205/765/1/L12
[arXiv:1302.3180 [astro-ph.SR]].

\bibitem{Panotopoulos:2018sso}
G.~Panotopoulos and \'A.~Rinc\'on,
Phys. Rev. D \textbf{97} (2018) no.10, 103509
doi:10.1103/PhysRevD.97.103509
[arXiv:1804.11208 [astro-ph.CO]].

\bibitem{Panotopoulos:2019xbw}
G.~Panotopoulos, \'A.~Rinc\'on, G.~Otalora and N.~Videla,
Eur. Phys. J. C \textbf{80} (2020) no.3, 286
doi:10.1140/epjc/s10052-020-7828-7
[arXiv:1912.01723 [gr-qc]].

\bibitem{Panotopoulos:2017mte}
G.~Panotopoulos,
Phys. Rev. D \textbf{96} (2017) no.2, 023520
doi:10.1103/PhysRevD.96.023520
[arXiv:1706.10211 [astro-ph.CO]].

\bibitem{Zimdahl:2003wg}
W.~Zimdahl and D.~Pavon,
Gen. Rel. Grav. \textbf{36} (2004), 1483-1491
doi:10.1023/B:GERG.0000022584.54115.9e
[arXiv:gr-qc/0311067 [gr-qc]].

\bibitem{Sahni:2002fz}
V.~Sahni, T.~D.~Saini, A.~A.~Starobinsky and U.~Alam,
JETP Lett. \textbf{77} (2003), 201-206
doi:10.1134/1.1574831
[arXiv:astro-ph/0201498 [astro-ph]].

\bibitem{Panotopoulos:2007zn}
G.~Panotopoulos,
Nucl. Phys. B \textbf{796} (2008), 66-76
doi:10.1016/j.nuclphysb.2007.12.001
[arXiv:0712.1177 [astro-ph]].

\bibitem{Will:2014kxa}
C.~M.~Will,
Living Rev. Rel. \textbf{17} (2014), 4
doi:10.12942/lrr-2014-4
[arXiv:1403.7377 [gr-qc]].

\end{thebibliography}
\end{document}